\documentclass[letterpaper,11pt]{article}

\def\thefootnote{\arabic{footnote}}

\usepackage{epsfig}
\usepackage{graphicx}% Include figure files
\DeclareMathAlphabet   {\mathsc}{OT1}{cmr}{m}{sc}

\def\[{\left [}
\def\]{\right ]}
\def\({\left (}
\def\){\right )}
\newcommand{\lang}{\left\langle}
\newcommand{\rang}{\right\rangle}
\newcommand{\lbr}{\left\{}
\newcommand{\rbr}{\right\}}
\newcommand{\beq}{\begin{equation}}
\newcommand{\eeq}{\end{equation}}
\newcommand{\bea}{\begin{eqnarray}}
\newcommand{\eea}{\end{eqnarray}}
\newcommand{\oline}[1]{\overline{#1}}

\newcommand{\wtd}[1]{\widetilde{#1}}
\newcommand{\GeV}      {~\mathrm{GeV}}
\newcommand{\TeV}      {~\mathrm{TeV}}

\newcommand{\order}{\mathcal{O}}

\newcommand{\gappeq}{\mathrel{\rlap {\raise.5ex\hbox{$>$}}
{\lower.5ex\hbox{$\sim$}}}}
\newcommand{\lappeq}{\mathrel{\rlap{\raise.5ex\hbox{$<$}}
{\lower.5ex\hbox{$\sim$}}}}

\newcommand{\DSab}{(\Delta S_{AB})^2}
\newcommand{\Ochi}{\Omega_{\chi} {\rm h}^2}

\begin{document}

%\begin{titlepage}
\begin{center}

\vskip .1in {\large \bf Solving the LHC Inverse Problem with Dark
Matter Observations}

\vskip .4in Baris~Altunkaynak, Michael~Holmes and Brent~D.~Nelson
\vskip .1in

{\em Department of Physics, Northeastern University, Boston, MA
02115} \vskip .1in

\end{center}

\begin{abstract}

    We investigate the utility of cosmological and astrophysical observations
    for distinguishing between supersymmetric theories. In particular we consider
    276~pairs of models that give rise to nearly identical patterns of observables at hadron
    colliders. We focus attention on neutralino scattering
    experiments (direct detection of relic neutralinos) and
    observations of gamma-rays from relic neutralino annihilation
    (indirect detection experiments). Both classes of experiments
    planned for the near future will make measurements with
    exceptional precision. In principle, therefore, they will have
    the ability to be surprisingly effective at discriminating between
    candidate theories. However, the ability to distinguish between
    models will be highly dependent on future theoretical progress
    on such things as determination of the local
    halo density model and uncertainty in nuclear matrix elements
    associated with neutralino recoil events. If one imagines
    perfect knowledge of these theoretical inputs, then with extremely conservative
    physics assumptions and background estimates we find 101~of the
    276~degenerate pairs can be distinguished.
    Using slightly more optimistic assumptions about background rates
    increases this number to~186 of the 276~pairs. We discuss the sensitivity of
    these results to additional assumptions made
    about nuclear matrix elements, the cosmological density of neutralinos and the galactic
    halo profile. We also comment on the complementarity of this study to recent
    work investigating these same pairs at a $\sqrt{s} = 500 \GeV$ linear collider.

\end{abstract}
%\end{titlepage}
%\newpage

%\renewcommand{\theequation}{\arabic{section}.\arabic{equation}}
\renewcommand{\thepage}{\arabic{page}}
\setcounter{page}{1}
\def\thefootnote{\arabic{footnote}}
\setcounter{footnote}{0}

\section{Introduction}

With the first data from the Large Hadron Collider (LHC) rapidly
approaching there is a shift in focus occurring in the theoretical
community from how observations can be {\em made} to how
observations can be {\em used}. That is, assuming physics beyond the
Standard Model (BSM) is discovered in the early stages of LHC data
collection, how can the large set of measurements that will
eventually be made be used to reconstruct the Lagrangian of the
underlying physics model? There are many approaches to answering
this question. The direct approach assumes a rough ansatz for the
BSM physics can be postulated early on, and then mass eigenstate
information can be extracted by reconstructing specific decay chains
within this model framework~\cite{reconstruct}. Another, more
top-down approach is to perform direct fits to certain theoretically
motivated models which are presumed to be defined with a relatively
small number of input parameters~\cite{fits}. Recently, several
groups have begun working on hybrid approaches that combine
bottom-up exclusive measurements with top-down global analyses, such
as the construction of on-shell ef\mbox{}fective
theories~\cite{ArkaniHamed:2007fw} or the identification of mass
hierarchy patterns~\cite{Feldman:2007zn}. The possible redundancy of
these techniques is a welcome cross-check, and all are likely to be
of utility in the early stages of LHC data-taking. This process of
reconstructing both the form and the parameters of the underlying
new physics Lagrangian is commonly referred to as ``inverting'' the
LHC data.

An obvious question to ask is whether this reconstruction procedure
will yield a unique solution. The answer is likely to be yes if a
relatively simple and well-constrained theory gives a good fit to
the data~\cite{Binetruy:2003cy}. But for even slightly more
complicated theories the answer becomes ambiguous. Recently,
Arkani-Hamed et al.~\cite{ArkaniHamed:2005px} studied the minimal
supersymmetric standard model (MSSM), with 15~free parameters used
to determine the resulting spectrum and phenomenology. The authors
found that it is highly probable (perhaps inevitable) that even
after hundreds of measurements -- consisting of counting events with
distinct final-state topologies as well as examining the shapes of a
wide range of kinematic distributions -- more than one set of these
15~parameters will be a good fit to the data. In fact, a set of
parameters which fit the data may well have {\em several} such
``degenerate twins'' in the parameter space of the theory. The
challenge of disentangling these degenerate pairs is the LHC Inverse
Problem.

While the conclusion of~\cite{ArkaniHamed:2005px} is sobering, it
might not seem especially surprising. Within a model system as
complex as the MSSM it is possible to engineer two theories with
similar collider signatures because some aspects of the model (such
as the wavefunction of the lightest stable neutralino, or the value
of the parameter $\tan\beta$) are not easily probed in a hadron
collider. While it is very dif\mbox{}ficult to construct a
degenerate twin for any particular supersymmetric model, given
enough sampling of the parameter space it is not dif\mbox{}ficult to
find pairs of points that are degenerate to within the experimental
error desired. For example, over 43,000~parameter points were
considered in~\cite{ArkaniHamed:2005px}, yielding 283~degenerate
pairs of parameter sets. Degeneracy here was measured with respect
to 1808~observations in 10~fb$^{-1}$ of LHC data, after crude cuts
were made to reduce the Standard Model backgrounds. One can argue
about how many of these 283~pairs of models could, in fact, be
distinguished if only the authors had used a different set of
observations, or considered more integrated luminosity, or used more
exclusive techniques such as decay-chain reconstruction in the
analysis. Undoubtedly many of these pairs would indeed be
distinguished, eventually, at the LHC. Additional research in each
of these directions (using the degenerate pairs
of~\cite{ArkaniHamed:2005px} as a guide) is warranted.

In this letter we choose to take the primary conclusion
of~\cite{ArkaniHamed:2005px} at face value. Thus we will assume that
(R-parity conserving) supersymmetry is discovered early at the LHC.
Yet we also assume that even as LHC data accumulates isolated
parameter sets for the MSSM emerge which cannot be distinguished
with LHC data alone. We will use the 283~degenerate pairs
of~\cite{ArkaniHamed:2005px} as proxies for these post-LHC
degeneracies. Here we will investigate the efficacy of cosmological
and astrophysical observations associated with the stable relic
neutralino in distinguishing between these pairs. We believe this to
be a particularly fruitful area to study. The conspiracy of soft
supersymmetry breaking parameters necessary for two models to give
similar signatures at the LHC often gave rise to spectra of gauginos
with common mass differences. The wave-functions of these mass
eigenstates, however, were often dramatically different between the
two degenerate models. Unfortunately, the wave-function composition
of the lightest neutralino (usually the lightest supersymmetric
particle, or LSP) is particularly difficult to measure at the LHC.
On the other hand, it is well known to have dramatic effects on the
cosmology of this stable LSP -- both in its thermal relic
abundance~\cite{Griest:1991gu} and on the prospects for its
experimental detection at dark matter related
experiments~\cite{BirkedalHansen:2002am}.

The present work is very much in the spirit of (and complementary
to) recent work~\cite{Berger:2007yu} in which these same pairs were
studied at a $\sqrt{s} = 500 \GeV$ international linear collider
(ILC). The results of that study were mixed. When charged
superpartners were kinematically accessible they were (more often
than not) also detectable above the Standard Model background. When
only neutral superpartners were kinematically accessible that was
not true. When one or both of the degenerate models in a pair had an
accessible and visible charged superpartner they were generally
distinguishable at a 500~GeV linear collider. Unfortunately only
57~pairs met this criterion with a $5\sigma$ level of
distinguishability (the number becomes only 63 when the criterion is
relaxed to the $3\sigma$ level). At issue is the relatively high
mass of low-lying gaugino states in the degenerate pair sample
of~\cite{ArkaniHamed:2005px}. The authors of~\cite{Berger:2007yu}
conclude that a center-of-mass energy of $\sqrt{s} = 1\TeV$ would
fare much better at distinguishing between these pairs. But given
the uncertainties currently surrounding the fate of any
international linear collider effort, it seems prudent to ask
whether additional information from outside the collider arena can
profitably be brought to bear on the issue of breaking degeneracies
within the allowed supersymmetric parameter space.

To that end, we will introduce the degenerate model pairs in
Section~\ref{sec:pairs}, briefly discussing the methodology of
Arkani-Hamed et al. in Section~\ref{sec:inverse}. A discussion of
our analysis of these pairs begins in Section~\ref{sec:constraints}
with a look at some of the global properties of this model set. We
will find that almost all of the models predict a thermal relic
abundance outside a 95\% confidence level region about the preferred
value deduced from observations of the cosmic microwave background
by the WMAP experiment. This is not a surprise since none were
designed to meet this constraint (or indeed several other indirect
constraints on supersymmetric models). This observation leads to a
natural classification of the models into those which are at least
{\em consistent} with all indirect constraints, and those which are
at odds with at least one such constraint. We will use this
classification as a guide to present our subsequent results. In
Section~\ref{sec:direct} we discuss the ability of direct detection
experiments to distinguish models, and in Section~\ref{sec:gamma} we
discuss the same for indirect detection experiments that look for
photons from relic neutralino annihilation. The net effect of all of
these experiments in differentiating between the degenerate models
is presented in the concluding section.

One final comment is in order before we begin. By
``distinguishable'' we here mean that any two predicted signals --
such as the number of observed nuclear recoils in a direct detection
experiment -- could be distinguished from one another with a high
degree of statistical significance. When theoretical uncertainties
are neglected the previous sentence becomes a statement about the
inherent resolving power of a particular experiment (or
experiments). For example, this is what is typically meant by the
concept of distinguishing between models in studies that focus on
collider signals. But the dark matter arena is in many ways more
difficult than studying signatures at the LHC or ILC: rather than
studying similar detectors at a single facility with a well-modeled
background (and high event rates) we must here work with a large
variety of experimental configurations. Event rates are generally
low and background estimations less well understood than in the
collider environment. Finally, it is generally the case that
theoretical inputs, such as the assumed dark matter halo profile or
the value of certain nuclear matrix elements, are often the largest
sources of uncertainty. Statements about distinguishing candidate
theories must always be understood within this context. Because of
these uncertainties separating two post-LHC candidate models will be
difficult without the assistance of having a detailed astrophysical
model {\em as well as} a concrete particle physics model (or in this
case, a pair of models) at hand. This point has been emphasized
recently in~\cite{Bourjaily:2005ax}. Therefore, in this paper we
will generally make statements in which a fixed astrophysics model
(such as a particular galactic halo profile) is assumed, as is
common in the literature. ``Distinguishability'' under these
assumptions then essentially reduces to a consistency check between
one of a pair of models suggested by the LHC data and an assumed
dark matter signal. Within this framework we make every effort to be
as conservative as possible and present our results under a variety
of assumptions so that the reader can make his or her own
comparisons. We will return to the issue of theoretical
uncertainties and the broader notion of separating models throughout
this work, and make comments on how they might be potentially
resolved in the conclusion.

%\pagebreak
%%%%%%%%%%%%%%%%%%%%%%%%%%%%%%%%%%%%%%%%%%%%%%%%%%%%%%%%%%%%%%%%%
%%% SECTION TWO
%%%%%%%%%%%%%%%%%%%%%%%%%%%%%%%%%%%%%%%%%%%%%%%%%%%%%%%%%%%%%%%%%
\section{The Degenerate Pairs}
\label{sec:pairs}

\subsection{Summary of Arkani-Hamed et al.}
\label{sec:inverse}

In this subsection we give a very brief summary of the methodology
of Arkani-Hamed et al.~\cite{ArkaniHamed:2005px}, which serves as a
way of introducing the degenerate model pairs. As mentioned in the
introduction, the theory considered was that of the MSSM, with
15~parameters used to define the spectrum and interactions
\beq \lbr \begin{array}{c} \tan\beta,\,\,\mu,\,\,M_1,\,\,M_2,\,\,M_3 \\
m_{Q_{1,2}},\,m_{U_{1,2}},\,m_{D_{1,2}},\,m_{L_{1,2}},\,m_{E_{1,2}} \\
m_{Q_3},\,m_{U_3},\,m_{D_3},\,m_{L_3},\,m_{E_3} \end{array} \rbr \,
. \label{paramset} \eeq
Here $\tan\beta$ is the ratio of the two scalar Higgs vevs, $\mu$ is
the supersymmetric Higgsino mass parameter and $M_i$ represent the
soft supersymmetry breaking masses of the gauginos. Note that all
three gaugino masses are independent here, allowing for a wide array
of predictions in the dark matter arena. In the next two lines
of~(\ref{paramset}) $m_{X_{1,2}}$ is the soft supersymmetry breaking
mass of the first two generations of the scalar field $X$, taken to
be identical, while $m_{X_{3}}$ is soft mass for the third
generation of field $X$. We will refer to any such set of 15~values
for the quantities in~(\ref{paramset}) as a ``model.'' Other
parameters required to establish the spectrum were taken as fixed,
and thus are the same for all models considered: the pseudoscalar
Higgs mass at $m_A = 850\GeV$ and the scalar trilinear couplings at
$A = 800\GeV$ for third generation squarks and $A=0$ for the
sleptons. These are parameters valid at the electroweak scale, so no
renormalization group evolution is required. Constraints were placed
on the sizes of some parameters in order to make the collider
simulation tractable.\footnote{In particular, soft parameters
associated with fields carrying $SU(3)$ charge were required to be
larger than $600 \GeV$ and all states were required to have mass
parameters larger than $100 \GeV$. This had a sizable impact on the
accessibility of these states at a $\sqrt{s} = 500 \GeV$ linear
collider~\cite{Berger:2007yu}.}

Over 43,000~parameter sets were chosen at random and 10~fb$^{-1}$ of
LHC data was generated for each such model -- a massive
computational undertaking. Data was simulated using {\tt
PYTHIA}~\cite{Sjostrand:2006za} and the detector response was
modeled using the {\tt PGS} detector simulator~\cite{PGS}. A set of
initial cuts were used to simulate what would be required in a true
analysis to reduce the Standard Model background, though no actual
Standard Model processes were included in the analysis itself. This
analysis had two components. The first counted the number of events
which had one of a number of suitably defined final state
topologies. The second component included studying the shapes of a
number of key kinematic distributions of final state decay products.
These shapes were parameterized by taking the relevant histogram and
binning the data into even numbers of quantiles, then recording the
position of the quantile boundaries. In this way both components of
the data analysis could be included in a $\chi^2$-like variable. In
total there were 1808~such quantities $s_i$ for each model, the set
of all~1808 being called the ``signature'' of the model at the~LHC.

These individual $s_i$ values were grouped together into a variable
similar to a traditional chi-squared quantity,
\beq (\Delta\,S_{AB})^2 = \frac{1}{N_{\rm sig}}\sum_i\(\frac{s_i^A -
s_i^B}{\sigma_i^{AB}}\)^2 \, , \label{DeltaS} \eeq
where $A$ and $B$ represent two different models, $N_{\rm sig}$ is
the total number of signatures considered and $\sigma_i^{AB}$ is a
measure of the error associated with the $i$-th signature
\beq \sigma_i^{AB} = \sqrt{(\delta s_i^A)^2 + (\delta s_i^B)^2 +
\(f_i \frac{s_i^A + s_i^B}{2}\)^2}\, . \label{sigmaAB} \eeq
The final quantity is meant to represent the error associated with
incomplete removal of Standard Model background events from the data
sample. The value $f_i = 0.01$ was chosen for all observables except
for the total event rate, for which $f_i = 0.15$ was chosen. The
quantity~(\ref{DeltaS}) therefore provides a reasonable metric for
measuring distances in signature space. The next question is how far
apart two models need to be in this space to be deemed
``distinguishable.'' There are a number of reasonable answers. The
criterion chosen in~\cite{ArkaniHamed:2005px} is as follows. Imagine
taking any supersymmetric theory and performing a collider
simulation. Now choose a new random number seed and repeat the
simulation. Due to random fluctuations we expect that even the same
set of input parameters, after simulation and event reconstruction,
will produce a slightly different set of signatures. Now repeat the
simulation a large number of times, each with a different random
number seed. Use~(\ref{DeltaS}) to compute the distance of each new
``model'' with the original simulation. While we might expect the
distribution of $\DSab$ to be narrow with a central value near zero,
we nevertheless expect there to be some spread. Find the value of
$\DSab$ which represents the 95th percentile of the distribution.
This might be taken as a measure of the uncertainty in ``distance''
measurements associated with statistical fluctuations.
In~\cite{ArkaniHamed:2005px} this number was determined to be
$\DSab\big|_{\rm 95th} = 0.285$. Any two genuinely different models
for which the calculated $\DSab$ is less than 0.285 were therefore
considered to be degenerate with each other.

Applying this criterion, Arkani-Hamed et al.~found 283~pairs of
models which failed to be distinguished at the LHC in their study
(albeit with only 10 fb$^{-1}$ of data). As many models were
degenerate with more than one other set of parameters, the total
number of unique individual parameter sets represented in this
sample was~384, from an original set of over 43,000 models.
%
%\footnote{\textbf{The ILC folks list 383 models and 283 pairs. Not
%sure why there is a difference...}}
%
We were generously provided with the values of the parameters
in~(\ref{paramset}) for these 384~models from the collaboration
in~\cite{ArkaniHamed:2005px}. We used these input parameters to
generate 50,000~events for each model at the LHC, roughly equivalent
to 5~fb$^{-1}$ of data, using the same combination of {\tt PYTHIA} +
{\tt PGS}. While a complete reproduction of the analysis in that
work would beyond the scope of our study, we did use the variables
defined in~(\ref{DeltaS}) and~(\ref{sigmaAB}) to examine a reduced
set of~36 signatures in the generated data. We confirmed that the
283~model pairs were highly degenerate at the LHC, most falling well
within our threshold $\DSab\big|_{\rm 95th} = 0.63$ for
indistinguishability.

\subsection{Classification and further constraints}
\label{sec:constraints}

Before proceeding to a study of dark matter observations, we wish to
consider a few global aspects of this model set. In studying the
signatures of these models at a future linear collider the authors
of~\cite{Berger:2007yu} encountered an initial problem: {\tt PYTHIA}
computes physical gaugino masses only at the tree-level. Many of the
models in this set have soft supersymmetry breaking gaugino masses
such that the lightest chargino is slightly less massive than the
lightest neutralino. This is normally not a disaster as loop
corrections to the mass eigenstates remedy the problem. In {\tt
PYTHIA}, however, the problem is instead rectified by artificially
adjusting the chargino mass to be the mass of the lightest
neutralino plus twice the neutral pion mass. In {\tt PYTHIA} 6.4
this comes with a warning flag. In our analysis 149~of the
384~models were so flagged. A similar number of problematic models
were observed in~\cite{Berger:2007yu}. In studying the linear
collider signatures of these models the mass difference between
these two states is a crucial parameter in determining the
detectability of the low-lying chargino state. Without a reliable
calculation of this mass difference the authors
of~\cite{Berger:2007yu} decided to jettison these models. In our
case the mass difference is crucial only in the determination of the
relic neutralino density, $\Ochi$, since the rate of
neutralino-chargino coannihilation in the early universe is very
sensitive to this quantity.

%=============== Relic density versus mass ===================
\begin{figure}[t]
\begin{center}
\includegraphics[scale=1.0,angle=0]{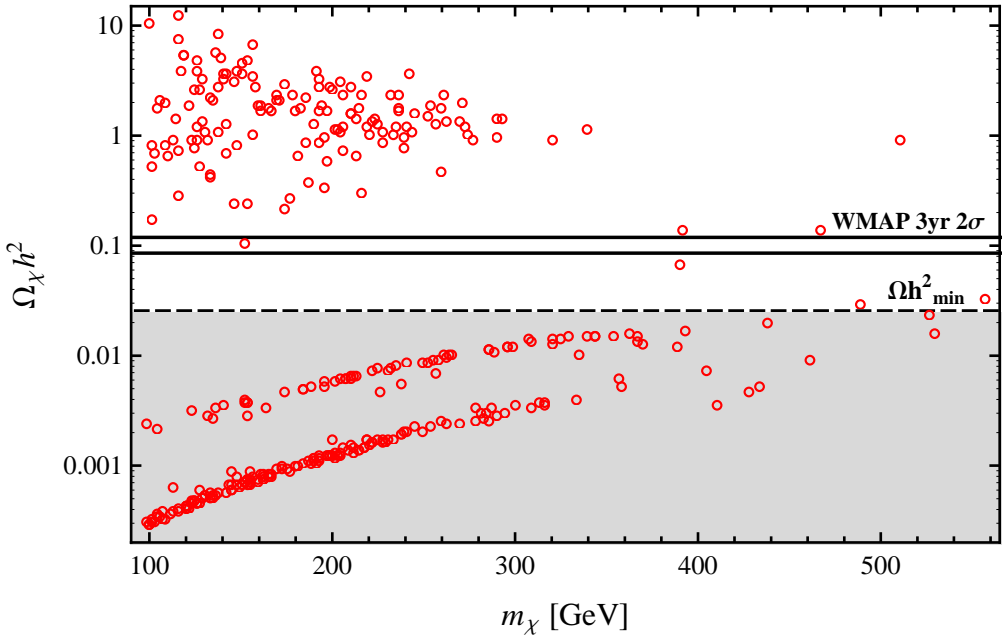}
\caption{\footnotesize \textbf{Thermal relic abundance of
neutralino~LSP for 378~models from~\cite{ArkaniHamed:2005px}.}
Prediction for the thermal relic abundance $\Ochi$, as computed by
{\tt DarkSUSY}, is displayed as a function of the LSP mass
$m_{\chi}$. The $2\sigma$ band in this quantity favored by the WMAP
three-year data set is indicated by the solid horizontal lines. The
shaded region bounded by the dashed line represents the set of
models for which the local number density of neutralinos should be
rescaled.} \label{fig:Omegavsmass}
\end{center}
\end{figure}
%===================================================================

We compute this relic density using the software package {\tt
DarkSUSY}~\cite{Gondolo:2004sc} which includes the most significant
one-loop corrections to the neutralino mass matrix, eliminating the
problem with the chargino/neutralino mass difference. We will
discuss the results of that computation below. Here we merely point
out that the cosmological density represented by the number $\Ochi$
is not directly related to any of the observables we will be using
to distinguish between models. Instead it is the local halo density
that is relevant and the relationship between the two quantities is
not a precise one. We will therefore keep all~149 of these model
points in our analysis that follows. However, for six models the
{\tt DarkSUSY} spectrum calculation returned a stau as the lightest
supersymmetric particle (LSP), which is certainly a problem for
computing cosmological observables. We will eliminate all six of
these models from our study, leaving us with 378 models comprising
276 degenerate pairs. The distribution of thermal relic neutralino
abundances for these 378 models is shown in
Figure~\ref{fig:Omegavsmass} as a function of the mass of the
lightest neutralino. The narrow band indicated by the solid
horizontal lines is the $2\sigma$ region favored by the WMAP
three-year data~\cite{Spergel:2006hy}
\beq 0.0855 < \Ochi < 0.1189\, . \label{omegah2} \eeq
All but one of the 378~models lie outside the band of values favored
by the WMAP data; 145~exceed the upper bound while 232~fall below
the lower value. Of the latter, 224 have $\Ochi \leq 0.025$
(indicated by the dashed line in Figure~\ref{fig:Omegavsmass}). This
value is used as a crude measure of the point at which the relic
particle in question can no longer account adequately for the local
halo density of our galaxy~\cite{Jungman:1995df}. We will return to
this issue in a moment.

The implication of Figure~\ref{fig:Omegavsmass} is that there are
{\em no} pairs among the~276 for which both models in the pair give
a relic abundance within the range of~(\ref{omegah2}). This should
not be interpreted to mean that degenerate pairs do not exist which
are fully consistent with this range -- nor that such degenerate
pairs are in particularly obscure parts of the supersymmetric
parameter space. It means nothing more than that the conventional
top-down method of searching for such pairs is horribly inefficient.
The authors of~\cite{ArkaniHamed:2005px} were interested in general
issues of LHC~phenomenology for which the value of $\Ochi$ is wholly
irrelevant. In fact, restricting their attention only to models
which satisfy~(\ref{omegah2}) would have unduly biased their study.
The ultimate relic density of neutralinos can be a very sensitive
function of the masses and mixings of the superpartner spectrum. We
have no doubt that many of these model pairs could be adjusted to
give degenerate results at the~LHC while providing for a reasonable
value of $\Ochi$. Therefore we will analyze the dark matter related
signatures for all 378~model points in what follows. Nevertheless we
will group those which exceed the upper limit in~(\ref{omegah2})
separately, as it is generally quite difficult to engineer processes
in the early universe to reduce the thermal abundance of a stable
relic~\cite{Gelmini:2006pw}. By contrast, it is not hard to imagine
ways to enhance the relic abundance of a stable neutralino through
non-thermal mechanisms~\cite{Moroi:1999zb}. For that reason we will
not eliminate models for which $\Ochi$ is below the lower bound
of~(\ref{omegah2}) or consider them unphysical. However, when
calculating observable quantities that depend on the relic
neutralino number density $n_{\chi}$ present in our galaxy (or the
energy density $\rho_{\chi} = m_{\chi} n_{\chi}$) we will rescale
the assumed local density of $(\rho_{\chi})_0 = 0.3$~GeV/cm$^3$ by
the multiplicative factor $r_{\chi} = {\rm Min}(1,\, \Ochi/0.025)$.
%for those models in which $\Ochi < 0.025$.

In~\cite{ArkaniHamed:2005px} the authors were careful to ensure that
all direct search bounds for superpartners were satisfied by the
resulting spectrum. They were less concerned with the bound on the
lightest CP-even Higgs mass from LEP, nor with other indirect
constraints on supersymmetry. Consider, for example, the following
bounds taken from~\cite{Djouadi:2006be}
\beq \begin{array}{c} m_h \geq 114.4 \GeV \\ 2.65 \times 10^{-4}
\leq {\rm Br}(B \to X_s \gamma) \leq 4.45 \times 10^{-4} \\
\frac{g_{\mu} -2}{2}\big|_{\rm SUSY} \leq 4.7 \times 10^{-9}
\end{array} \label{constraints} \eeq
for the mass of the lightest CP-even Higgs, branching fraction for
$b \to s \gamma$ events, and the supersymmetry contribution to the
anomalous magnetic moment of the muon, respectively. These bounds
were violated by 43, 101 and 6 models, respectively, from the set
of~378. All of these have marginal relevance to the physics of
the~LHC, and even less relevance to the physics we discuss in this
paper. Nevertheless, we will designate a subset of the 378~models as
``physical'' if they have a relic density below the upper bound
in~(\ref{omegah2}) and satisfy all three
constraints~(\ref{constraints}). There are then 127~such
``physical'' models, arranged in 77~indistinguishable pairs.
Throughout the subsequent work we will focus on this physical subset
where appropriate.

%---------------------- Summary Table -------------------------
\begin{table}[t]
\begin{center}
\begin{tabular}{|l||c|c|}
\multicolumn{1}{c}{}
 & \multicolumn{1}{c}{Models}
 & \multicolumn{1}{c}{Pairs} \\ \hline
Initial Set & 378 & 276 \\
\mbox{         } {\tt PYTHIA} chargino warnings & 149 & 124 \\
\hline
Relic Density & & \\
\mbox{         }   $\Ochi > 0.1189$ & 145 & 116 \\
\mbox{         }   $\Ochi < 0.0250$ & 224 & 164 \\ \hline
Additional Constraints & & \\
\mbox{         } $m_h < 114.4$ GeV & 43 & 52 \\
\mbox{         } ${\rm Br}(B \to X_s \gamma) > 4.45 \times 10^{-4}$ & 101 & 98 \\
\mbox{         } $a_{\mu}\big|_{\rm SUSY} > 4.7 \times 10^{-9}$ & 6
& 6 \\ \hline
Visible at 500~GeV ILC & 190 & 173  \\
\mbox{         } Remove {\tt PYTHIA} chargino warnings & 68 & 65 \\
\hline
All Physical Conditions Satisfied & 127 & 77  \\
\hline
\end{tabular}
\end{center}
{\caption{\label{summary}\footnotesize {\bf Summary of global
properties of the 378~models}. Statistics listed under the heading
``Models'' give the number of individual model points satisfying the
stated property. Statistics in the column labeled ``Pairs'' count
the number of pairs where {\em at least one} of the two models in
the pair satisfies the stated property. If a model has $\Ochi <
0.1189$ and fails none of the tests listed in the Additional
Constraints section it is counted in the last row of the table. If
{\em both} models in a pair do so, the {\em pair} is counted in the
last row.}}
\end{table}
%------------------------- END OF THE TABLE ---------------------

Finally, it is useful to consider how dark matter based observations
can buttress the results of~\cite{Berger:2007yu}. While we do not
know precisely which of the 276~pairs were deemed to be
distinguishable at a $\sqrt{s} = 500 \GeV$ linear collider, we can
estimate which models would be visible at such a machine by simply
requiring that a chargino or charged slepton be kinematically
accessible. To be conservative we require a mass less than $240
\GeV$ to be deemed kinematically accessible. With this assumption we
estimate that 190~models would be visible at a 500 GeV linear
collider; or only 68~models if we reject those generating a chargino
mass warning flag in {\tt PYTHIA}. This is to be compared with
63~model pairs that were found to be distinguishable at $3\sigma$
in~\cite{Berger:2007yu}, where at least one model was accessible and
detectable. We will comment on this subdivision of the 378~models in
the conclusion section of this work. A summary of the various
classifications introduced in this section is given in
Table~\ref{summary}.

%%%%%%%%%%%%%%%%%%%%%%%%%%%%%%%%%%%%%%%%%%%%%%%%%%%%%%%%%%%%%%%%%
%%% SECTION THREE
%%%%%%%%%%%%%%%%%%%%%%%%%%%%%%%%%%%%%%%%%%%%%%%%%%%%%%%%%%%%%%%%%
\section{Direct Detection Experiments}
\label{sec:direct}

Before taking our first look at potential experimental data let us
recall the assumptions underlying our thought experiment. We imagine
that LHC data equivalent to several years running has been
accumulated and analyzed. This analysis has been restricted to
broadly inclusive observables of the sort considered by Arkani-Hamed
et al. -- that is, we do not allow ourselves knowledge of the mass
of individual mass eigenstates. Undoubtedly such information will
become available through the study of isolated decay chains and
kinematic end-point variables associated with them. But to be true
to the spirit of ``model-independence'' of~\cite{ArkaniHamed:2005px}
we do not allow any such knowledge in what follows. This is
particularly relevant in the case of the mass of the lightest
supersymmetric particle, upon which many observable quantities we
will consider depend. We will comment further on the power of this
information in the concluding section of this paper.

However we do imagine that some sort of global fit has been
performed, in which the values of these inclusive signatures have
been calculated as a function of a sufficiently broad parameter
space within the MSSM. At the end of this process two parameter sets
of the form of~(\ref{paramset}) ({\em i.e.} two ``models'') have
emerged as equally good fits to the observed data.\footnote{For an
example of what types of low-energy fits are possible,
see~\cite{Lafaye:2007vs}.} From this perspective our 276~degenerate
model pairs represent 276~different possible outcomes of this global
fitting process. We therefore have two candidate parameter sets
``A'' and ``B'' from which we can calculate the theoretical
expectation for various dark matter observables $s_i$ for both
models in the pair. To claim that a particular experiment has the
power to distinguish between models~A and~B we require two
properties simultaneously. First, the values of $s_i^A$ and $s_i^B$
need to be large enough that both are detectable above the relevant
background for the experiment in question. Second, the values of
$s_i^A$ and $s_i^B$ need to be sufficiently separated to give a
statistically significant difference when measured with respect to
the appropriate mutual error $\sigma_i^{AB}$. It is possible to
relax the first assumption by requiring only that one of the two
experiments yield a detectable signal $s_i$, which can be
distinguished statistically from $s_i = 0$. This was the approach
taken in~\cite{Berger:2007yu}. But given the inherent uncertainties,
both experimental and theoretical, associated with dark matter
observables we prefer a more conservative requirement for
distinguishability.

With this in mind we will focus in this section on direct detection
of relic neutralinos via their scattering from target nuclei.
Scattering events are signaled by the detection of the nuclear
recoil for elastic scatters, or by detecting the resulting
ionization of the target nucleus for inelastic
scattering~\cite{Gaitskell:2004gd}. We consider a range of current
and future experiments listed in Table~\ref{experiments}. For each
experiment we give the target nucleus, the target mass and the
physics object (or objects) which are actually detected to signal a
scattering event. The first three experiments in
Table~\ref{experiments} are currently taking data and setting limits
on nucleon-neutralino interaction cross sections. The target mass
for these experiments represents the true fiducial mass. The
remaining experiments are in various stages of development and
planning. The entry in the target mass column is only a nominal mass
-- the actual fiducial mass used for data taking is typically
smaller by a significant factor.\footnote{Where necessary in what
follows we will assume the fiducial mass is 80\% of the nominal
mass.}

%---------------------- Summary Table -------------------------
\begin{table}[t]
\begin{center}
\begin{tabular}{|c|l||c|c|l|} \hline
Ref. & Experiment Name & Target & Mass (kg) & Detected Object(s) \\
\hline
\cite{Ahmed:2008eu} & CDMS~II & Ge & 3.75& athermal phonons, ionization charge\\
% Integration: 10-100 keV
% Nominal mass: 4.75 kg
% Exposure: 397.8 kg-days
% Events seen: 0
% True fakes expected: <0.2 events total
% Fake fakes expected: less than one event (0.6 +/- 0.5 events)
%
\cite{Angle:2007uj} & XENON10 & Xe & 5.4& scintillation photons, ionization charge\\
% Integration: 4.5 - 26.9 keV
% Nominal mass: 15kg
% Exposure: 58.6days = 316.4 kg-days
% Events seen: 10
% True fakes expected: ?
% Fake fakes expected: 7 +/- 1.4/1.0 events
%
\cite{Alner:2007ja} & ZEPLIN~II  & Xe & 7.2& scintillation photons,
ionization charge \\ \hline \hline
% Integration: 5-20 keVee and keVnr = 2.8*keVee so this is 14-56 keV
% Nominal mass: 31kg
% Exposure: 225 kg-days
% Events seen: 29 (or 0.092 events/kg/day)
% True fakes expected: (in 25-50 keV recoil range for 30kg)
%                      20-30 events/year,
%                      but certainly less than 40 events/year,
%                      or less than 0.01 events/kg-day
% Fake fakes expected: (combined: 28.6 +/- 4.3 events)
%
\cite{Akerib:2006rr} & SuperCDMS (Soudan)  & Ge & 7.5& see CDMS~II\\
% Start time: Now
% Integration:
% Nominal mass: 7.5kg
% True fakes expected: (0.2) events in 2800 kg-day
% Fake fakes expected: (0.11) events in 2800 kg-day
%
\cite{Akerib:2006rr} & SuperCDMS (SNOlab) & Ge & 27& see CDMS~II\\
% Start time: 2010-2012 (LUX claims 2013)
% Integration: 15-45
% Nominal mass: 27kg
% True fakes expected: (0.004) events in 18000 kg-day
% Fake fakes expected: (0.6) events in 18000 kg-days
%
\cite{Akerib:2006rr} & SuperCDMS (DUSEL) & Ge & 1140& see CDMS~II\\
% Start time: after 2012, probably 2013
% Integration:
% Nominal mass: 1140 kg
% True fakes expected: (0.2) events in 10^6 kg-day
% Fake fakes expected: 33 events in 10^6 kg-day
%
\cite{Sanglard:2005we} & EDELWEISS-2 & Ge & 9& athermal phonons, ionization charge \\
% Start date: 2009/2010 (for low background runs)
% Integration: 10-100 keV (maybe just 30-100)
% Nominal mass: 9kg
% True fakes expected: 10^{-3}events/kg-day
% Fake fakes expected: 0.6 events/kg-day
%
\cite{Aprile:2004ey} & XENON100  & Xe & 170& see XENON10 \\
% Start time: Now
% Integration: 5keV+ (16keV+ in one paper)
% Nominal mass: 170kg
% Expectation: 100 events/year
% True fakes expected: 0.9 events/year (2.74 micro-dru)
% Fake fakes expected: 9.52 +/- 0.67 mdru = (10^{-3}x events/kg-day-keV)
%
\cite{Aprile:2004ey} & XENON1T  & Xe & 1000& see XENON10 \\
% Start time: 2009-2011
% Integration: 5keV+
% Nominal mass: 1000
% Expectation: 10 events/year
% True fakes expected:
% Fake fakes expected:
%
\cite{LUX} & LUX  & Xe & 350& scintillation photons, ionization charge \\
% Start time: Now-ish
% Integration: 5-25 keV
% Nominal mass: 350kg
% Fiducial volume: 100 kg
% Expectation: about 5 events per year
% True fakes expected: 3.2 events in 30,000 kg-days
% Fake fakes expected: 8x10^{-4} dru (1 event in 30,000 kg-days)
%
%& XMASS  & Xe & & \\
% Integration:
% Nominal mass:
% True fakes expected:
% Fake fakes expected:
%
\cite{Araujo:2006er} & ZEPLIN~III & Xe & 8& see ZEPLIN~II \\
% Start time: now (?)
% Integration: 5-50 keV
% Nominal mass:
% Fiducial mass: 8kg
% True fakes expected: ~10^{-3} dru (25 events/year in 8kg)
% Fake fakes expected: 10 dru
% Eventual goal after all upgrades: 9 neutrons and 2 gammas/year
%
\cite{Araujo:2006er} & ZEPLIN~IV  & Xe & 1000& see ZEPLIN~II \\
% Start time:
% Integration:
% Nominal mass:
% True fakes expected:
% Fake fakes expected:
%
\hline
\end{tabular}
\end{center}
{\caption{\label{experiments}\footnotesize {\bf List of direct
detection experiments considered}. The first three experiments
listed (CDMS~II, XENON10 and ZEPLIN~II) are currently taking data
and have recently reported limits on neutralino-nucleus interaction
rates. The masses for these three experiments are their respective
reported fiducial masses. All other experiments listed are projected
for some time in the future and the masses given are nominal mass
values.}}
\end{table}
%------------------------- END OF THE TABLE ---------------------

A couple of statements about this list are in order. These are not
all the experiments that one might consider, even if one restricts
attention to solely those based on germanium or xenon. The choice
here reflects the desire for simplicity of presentation and
reliability of background estimations. There are essentially two
classes of detector here: cryogenic germanium bolometers and
dual-phase liquid/gas xenon detectors. Both types are currently
operational in at least one experiment and producing results. Both
detector technologies have been shown to be scalable and
significantly larger installations of each technology are planned
(the nine experiments below the double line in
Table~\ref{experiments}). By taking measured background rates in
current experiments it is possible to extrapolate reliably to the
large scale experiments imagined in the future. Furthermore, by
focusing on two target materials it is possible to present the reach
and resolving power of multiple experiments in terms of just two
quantities: exposure time in germanium or xenon. This provides a
desirable degree of simplicity in presenting the results that
follow.

To compute the interaction rate of relic neutralinos with the nuclei
of the target material one considers both spin-dependent (SD) and
spin-independent (SI) interactions. For target nuclei with large
atomic numbers the SI interaction, which is coherent across all of
the nucleons in the nucleus, tends to dominate. This is true of
xenon and, to a slightly lesser extent, germanium as well. The SI
cross section $\sigma^{\mathrm{SI}}$ is computed in {\tt DarkSUSY}
on an arbitrary nuclear target via~\cite{Gondolo:2004sc}
\begin{equation}\label{ds30}
    \sigma^{\mathrm{SI}}_{\chi i}=\frac{\mu^2_{i\chi}}{\pi}\big|
    ZG^p_s+(A-Z)G^n_s \big|^2\, ,
\end{equation}
where $i$ labels the nuclear species in the detector with nuclear
mass $M_i$, $\mu_{i\chi}$ is the reduced mass of the
nucleus/neutralino system $\mu_{i\chi}=m_{\chi} M_i/(m_\chi + M_i)$,
and $A$ and $Z$ are the target nucleus mass number and atomic
number, respectively. The quantities $G^p_s$ and $G^n_s$ represent
scalar four-fermion couplings of the neutralino to point-like
protons and neutrons. They can be described schematically as
\begin{equation}
G^N_s=\sum_{q=u,d,s,c,b,t}\langle N|\bar q q | N\rangle \times \(
{\rm SUSY\,\,parameters} \)\, , \end{equation}
where the quantity in parenthesis is calculable once the details of
the supersymmetric model are specified. The initial nuclear matrix
elements, however, are at present not calculable from first
principles. Their values must be inferred from pion-nucleon
scattering data. Depending on the methodology employed in this
analysis, different values for this important set of parameters can
be extracted -- particularly for the case of the $\pi\,N$
$\Sigma$-term~\cite{Pavan:2001wz}. The importance of the resulting
uncertainty in this parameter on predictions for dark matter
interaction cross-sections was recently considered
in~\cite{Ellis:2005mb,Ellis:2008hf}, where it was shown to be
potentially quite large. We will return to this issue at the very
end of this section. For what follows we will simply use the default
values in {\tt DarkSUSY} for all nuclear matrix elements.

%=============== Cross section versus mass ===================
\begin{figure}[p]
\begin{center}
\includegraphics[scale=1,angle=0]{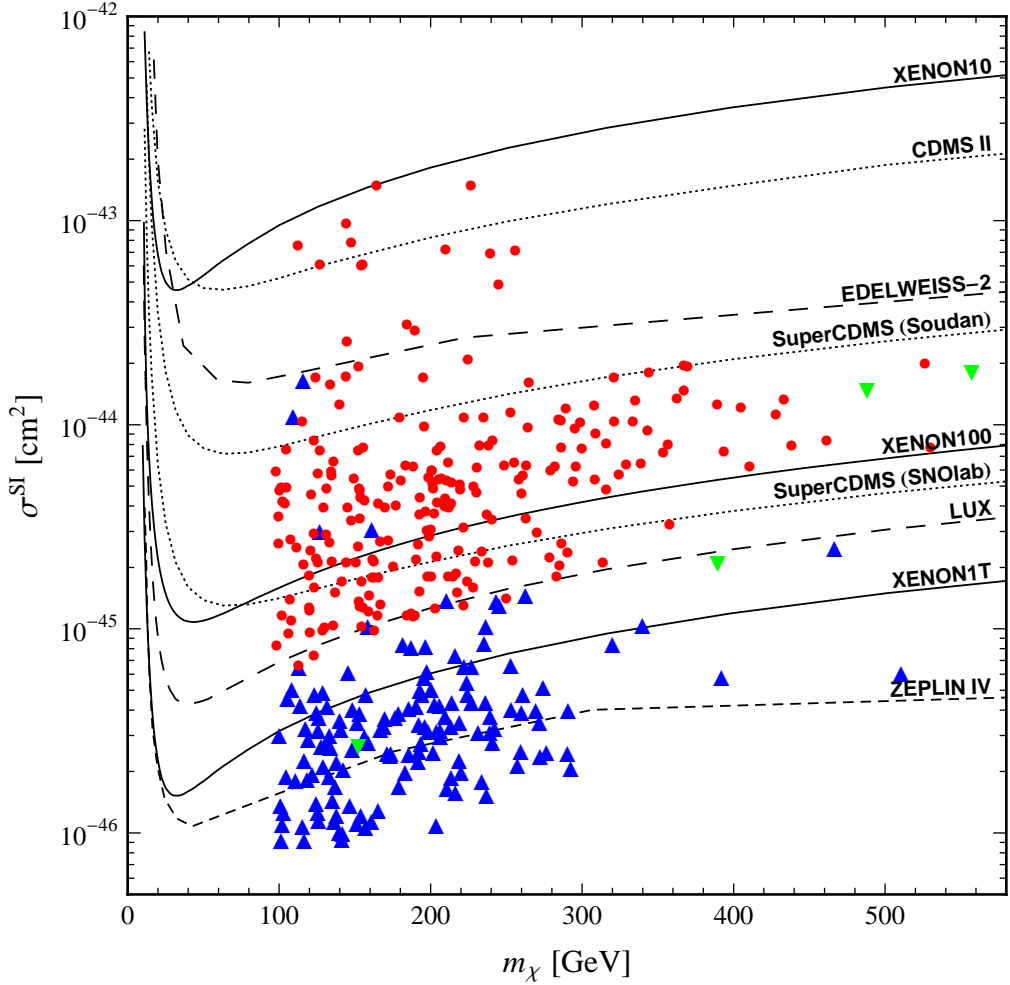}
\caption{\footnotesize \textbf{Spin independent neutralino-proton
interaction cross-section as a function of $m_{\chi}$ for the
378~models.} The 378~models are divided into three groupings: those
with $\Ochi > 0.1189$ (darker filled triangles), $ 0.025 < \Ochi <
0.1189$ (lighter inverted triangles) and $\Ochi < 0.025$ (filled
circles). Sensitivity curves for several of the experiments in
Table~\ref{experiments} are overlaid on the plot. These curves were
taken from the web-based utility~\cite{plotter}.}
\label{fig:Sigmavsmass}
\end{center}
\end{figure}
%===================================================================

It is common in the literature to graphically exhibit the reach of
any given experiment in terms of a variable in which results can be
compared easily. This {\em lingua franca} is
$\sigma^{\mathrm{SI}}_{\chi p}$ -- the interaction cross section of
the neutralino with the target nucleus, normalized to an equivalent
interaction cross section on protons~\cite{Kurylov:2003ra}. In
Figure~\ref{fig:Sigmavsmass} we therefore plot all 378~models in the
$(m_{\chi}\,,\sigma^{\mathrm{SI}}_{\chi p})$ plane. The points in
Figure~\ref{fig:Sigmavsmass} are separated into those for which
$\Ochi > 0.1189$ (darker filled triangles), $ 0.025 < \Ochi <
0.1189$ (lighter inverted triangles) and $\Ochi < 0.025$ (filled
circles). The nominal reach for a selection of the experiments in
Table~\ref{experiments} is also given, indicated by the various
lines as labeled in the figure. For the XENON10 and CDMS~II
experiment these lines represent actual exclusion curves. The limits
arising from the ZEPLIN~II data are weaker (on the order of
$10^{-42}$~cm$^2$) and therefore do not appear on the figure.

At first glance Figure~\ref{fig:Sigmavsmass} seems to indicate that
several of the models should have already given a discernable signal
in one or more current direct detection experiments. We would like
to argue that Figure~\ref{fig:Sigmavsmass} is somewhat deceptive in
this regard. Plots similar to Figure~\ref{fig:Sigmavsmass} are
typical in the high energy physics literature. They are sufficient
for a crude analysis of whether a particular experiment has the
sensitivity to ``discover'' or exclude a particular model in
question. But such a plot fails to adequately describe the challenge
of distinguishing two candidate theories. Experiments measure
counting rates, not cross sections. The relation between the two
involves additional assumptions about the local density of relic
neutralinos and their velocity distribution. As mentioned in
Section~\ref{sec:constraints} nearly 60\% of the models we are
studying have $\Ochi \leq 0.025$ -- including all of the models
whose cross sections are nominally above the XENON10 and CDMS~II
reach curves in Figure~\ref{fig:Sigmavsmass}. The experimentally
measured quantity -- the rate of nuclear recoils -- involves the
product of the spin-independent cross section and the neutralino
number density $n_{\chi}$. If we rescale this number density by
$r_{\chi} = {\rm Min}(1,\, \Ochi/0.025)$ we find that {\em none} of
the 378~models would yield more than one or two events in either
experiment in the reported exposure time accumulated thus far. This
exemplifies the importance of working directly with count rates, as
well as the importance of knowing the background rate for these
experiments.

An estimate of the rate of neutralino-nucleon scattering events (in
units of events/kg/day) can be computed from~(\ref{ds30}) according
to
\begin{equation}\label{rate est}
R\sim \sum_i\Phi_\chi \frac{\sigma^{\mathrm{SI}}_{\chi i}}{M_i} =
\sum_i \frac{\langle v_\chi
\rangle\rho_\chi\sigma^{\mathrm{SI}}_{\chi i}} {m_\chi M_i} \, ,
\end{equation}
where $\Phi_\chi$ is the average neutralino flux through the
detector and $\langle v_\chi \rangle$ is the average speed of the
neutralino relative to the target~\cite{Jungman:1995df}. For the
sake of a quick estimate one can take this velocity to be $\langle
v_\chi \rangle$ = 270 km/s.

For our rate calculations we use a more precise formulation that
takes into account the nature of the target. We begin by using {\tt
DarkSUSY} to compute the differential rate of interactions per unit
recoil energy via~\cite{Jungman:1995df}
\begin{equation}\label{ds29}
\frac{dR}{dE}=\sum_i c_i \frac{\rho_\chi \sigma_{\chi
i}|F_i(q_i)|^2} {2 m_\chi \mu^2_{i
\chi}}\int_{v_{min}}^\infty\frac{f(\vec v,t)}{v}d^3v \, .
\end{equation}
Now we sum over all nuclear species present, with $c_i$ being the
mass fraction of species $i$ in the detector. The quantity $f(\vec
v,t)\,d^3v$ is the neutralino velocity distribution (presumed to be
Maxwellian) with $v=|\vec v|$ the neutralino velocity relative to
the detector. Finally $|F_i(q_i)|^2$ is the nuclear form factor for
species $i$, with $q_i=\sqrt{2M_i E}$ being the momentum transfer
for a nuclear recoil with energy $E$. For the purpose of this
analysis we will use the output differential rates from {\tt
DarkSUSY}, calculated via~(\ref{ds29}), over a range of recoil
energies relevant to the desired experiment. For a given experiment
there is typically a minimum resolvable recoil energy $E_{\rm min}$
as well as a maximum recoil energy $E_{\rm max}$ that is considered.
These energies are $\order(10-100)$~keV and represent the nuclear
recoil energy of~(\ref{ds29}) inferred from the observed energy of
the detected physics objects. The range of integration is generally
different for each experiment and is determined by the physics of
the detector as well as the desire to maximize signal significance
over background. For example, the first three (active) experiments
in Table~\ref{experiments} integrate over the ranges
\bea {\rm CDMS\,\,II} &: & 10\,{\rm keV} \leq E_{\rm recoil} \leq
100\,{\rm keV} \nonumber \\
{\rm XENON10} &: & 4.5\,{\rm keV} \leq E_{\rm recoil} \leq
26.9\,{\rm keV} \nonumber \\
{\rm ZEPLIN\,\,II} &: & 14\,{\rm keV} \leq E_{\rm recoil} \leq
56\,{\rm keV}\, . \label{ranges} \eea

We perform a numerical integration of~(\ref{ds29})
%\begin{equation}\label{ratecalc}
%    R=\int_{E_{\rm min}}^{E_{\rm max}} \frac{dR}{dE}\,dE \, ,
%\end{equation}
%%
by constructing an interpolating function for the differential rate
sampled in $0.5$~keV intervals. Given the wide variety of
integration ranges exemplified by~(\ref{ranges}), and in order to
make meaningful comparisons across different experiments (including
the many experiments in Table~\ref{experiments} that are still in
the planning stages), we perform the integration using two possible
energy ranges
\bea R_1 &:& 5\,{\rm keV} \leq E_{\rm recoil} \leq 25\,{\rm keV}
\nonumber \\ R_2 &:& 10\,{\rm keV} \leq E_{\rm recoil} \leq
100\,{\rm keV} \, . \label{rates} \eea
The rate $R_1$ would cover the region that appears to be typical of
the dual-phase xenon detectors, while the larger range for the rate
$R_2$ seems to be typical for the germanium-based bolometer
experiments.  Empirically we find that the rate integrated over
$R_2$ is roughly twice that integrated over $R_1$:
\beq R^{\rm Xe}_2 = 1.86\, R^{\rm Xe}_1 \quad\quad\&\quad\quad
R^{\rm Ge}_2 = 2.24\, R^{\rm Ge}_1 \, . \eeq

Finally, to reverse-engineer the reach curves we must have some
notion of how well a particular experiment can distinguish nuclear
recoils due to neutralino scattering from fakes and background
events. This is also relevant to the question of whether two
possible signals can reliably be distinguished at any given
experiment. All of the experiments in Table~\ref{experiments}
collect charge as part of the detection process, and it is this
ionization charge that plays an important role in background
discrimination. Therefore background sources are quite similar
across the various types of experiments. We can break these
backgrounds into two crude classes: ``true'' neutron recoils and
``fake'' neutron recoils. In the former case we are thinking of
nuclear recoils that are measured in the detector but which did not
originate from a passing neutralino. They are generally the result
of neutrons produced in one of two ways: alpha-decays originating in
the material making up (or surrounding) the experimental chamber or
neutron recoils induced from cosmic-ray muons penetrating the
experimental chamber or surrounding materials. The fake neutron
recoils are cases where electric charge is collected in the
appropriate time window relative to the other physics object
(phonons or scintillation light), but where the electric charge is
induced by something other than a neutron recoil. This charge is
often caused by residual radioactivity in the detector elements,
especially the photomultiplier tubes that are present in many of
these experiments.  With proper shielding and a sufficiently
subterranean experiment site, the backgrounds from actual nuclear
recoils can be reduced to near zero. The background from electron
recoils is more difficult to eliminate -- further improvements in
background rejection will be necessary as the current experiments
scale to the one-ton limit. The sensitivity curves for future
experiments in Figure~\ref{fig:Sigmavsmass}, which we have taken
from~\cite{plotter}, already factor in some guess as to what these
improvements might be.

We would like to be able to discuss the entire collection of future
experiments as an ensemble when determining how many degenerate
pairs can be resolved. We therefore need some overall background
figure which we can apply universally (or perhaps one each for
germanium and xenon). Projections for large scale germanium-based
detectors are for background event rates of no more than a few
events per year of exposure. The liquid xenon detectors project a
slightly higher rate, but still on the order of 10-20 events per
year of exposure (mostly of the electron recoil variety). To be
conservative, therefore, we will make the following requirements on
two potentials signals to proclaim them distinguishable:
\begin{enumerate}
\item The count rates for the two experiments ($N_A$ and $N_B$),
obtained from integrating~(\ref{ds29}) over the appropriate range
in~(\ref{rates}), must {\em both} exceed $N$~events when integrated
over the exposure time considered. We will usually consider $N=100$,
but also show results for the weaker condition $N=10$.
\item The two quantities $N_A$ and $N_B$ must differ by at least
$n\,\sigma^{AB}$, where we will generally take $n=5$.
\end{enumerate}
We compute $\sigma^{AB}$ in a manner similar to~(\ref{sigmaAB}) by
assuming that the statistical errors associated with the measurement
are purely $\sqrt{N}$
\beq \sigma^{AB} = \sqrt{(1+f)(N_A + N_B)} \, , \label{oursigma}
\eeq
and the overall multiplicative factor $(1 + f)$ allows us to be even
more conservative by taking into account a nominal background rate
or allow for uncertainties in the local halo
density.\footnote{Please note the difference in form for this
additional ``fudge'' factor between~(\ref{oursigma})
and~(\ref{sigmaAB}). Also note that we are not yet considering the
issue of theoretical errors associated with the nuclear matrix
element uncertainty.} The case $f=0$ would therefore represent the
case of no background events.

%==================================================================
\begin{figure}[t]
%    \begin{center}
\centerline{
       \psfig{file=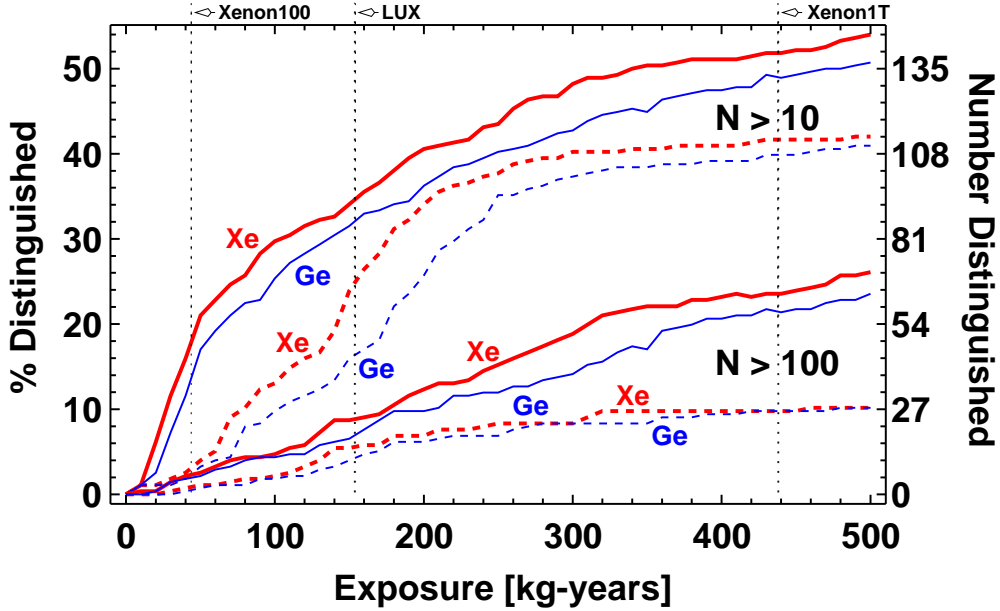,width=0.8\textwidth}
      }
          \caption{{\footnotesize {\bf Distinguishability analysis in 500 kg-years}.
          The number of degenerate pairs/percentage of the total that can be distinguished
          as a function of integrated exposure time. We plot exposure on germanium and
          xenon targets simultaneously. Heavy (red) lines are labeled for xenon,
          thinner (blue) lines are labeled for germanium. Solid lines have not been rescaled
          by the relic density ratio $r_{\chi}$, dashed lines have.
          The upper four lines are obtained by requiring only $N \geq 10$ recoil events for
          both models. The lower four lines are obtained by requiring $N\geq 100$ recoil
          events for both models. The predicted exposure after one year for three projected
          liquid xenon experiments is indicated by the vertical lines as labeled. Note that
          we assume 200 days of data-taking per calendar year with 80\% of the mass from
          Table~\ref{experiments} used as a fiducial target mass.}}
        \label{fig:ddsmall}
%    \end{center}
\end{figure}
%======================================================================

Based on these two criteria, none of the 378~models would have been
distinguished already in the Zeplin~II, CDMS~II or Xenon10
experiments -- indeed none should have produced a detectable signal
in any of these experiments. We do find nine models which would have
given at least ten events in 316.4 kg-days of exposure time in the
Xenon10 experiment, and five that would have given at least ten
events in 397.8 kg-days of exposure time in the CDMS~II experiment.
These are models that could have been discovered at CDMS~II (where
no signal-like events were observed) or nearly discovered at Xenon10
(where ten signal-like events were reported). Yet all of these cases
were ones in which the neutralino relic abundance was well below the
value $\Ochi = 0.025$. After rescaling the value of $\rho_{\chi}$
from its nominal value of $\rho_{\chi} = 0.3$ GeV/cm$^3$ (or
equivalently, the result of (\ref{ds29}) for the differential rate)
all of these models would have produced no events at either
experiment. This is despite the fact that the naive expectation from
Figure~\ref{fig:Sigmavsmass} is that at least a half-dozen models
would have been detected at CDMS~II, and at least one at Xenon10.

%==================================================================
\begin{figure}[t]
%    \begin{center}
\centerline{
       \psfig{file=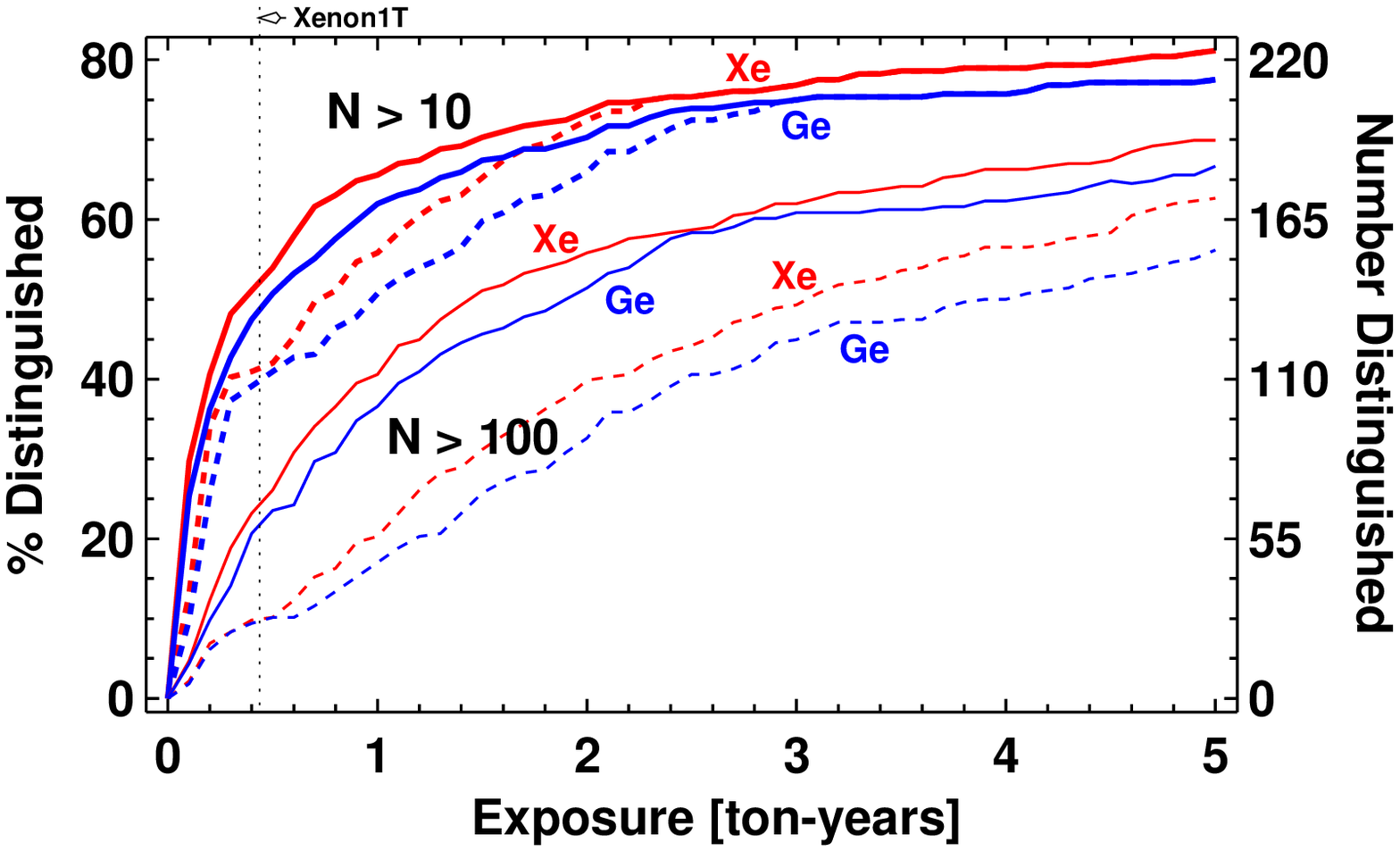,width=0.8\textwidth}
      }
          \caption{{\footnotesize {\bf Distinguishability analysis in 5~ton-years}.
          Same as Figure~\ref{fig:ddsmall} for a much larger exposure time. The vertical
          line represents our projection for one year of data-taking in the future
          XENON1T experiment. Note that
          we assume 200 days of data-taking per calendar year with 80\% of the mass from
          Table~\ref{experiments} used as a fiducial target mass.}}
        \label{fig:ddbig}
%    \end{center}
\end{figure}
%======================================================================

In Figures~\ref{fig:ddsmall} and~\ref{fig:ddbig} we plot the
percentage of the 276~pairs that can be distinguished as exposure
time is accumulated in xenon and germanium. Exposure time in
Figure~\ref{fig:ddsmall} is valued in units of kg $\times$
calendar-years and in units of tons $\times$ calendar-years in
Figure~\ref{fig:ddbig}. The separability criterion was $5\sigma$ and
assumed an experimental error with no additional smearing ($f=0$)
and no theoretical uncertainty. The upper four curves in the figures
require $N \geq 10$ predicted recoil events for each model before
being included in the total; the lower four curves require $N \geq
100$ predicted recoil events for each model. Solid lines are cases
in which the local halo number density $n_{\chi}$ was {\em not}
rescaled for models with $\Ochi < 0.025$. The dashed lines rescale
the local density (and hence the differential interaction rate) by
the parameter $r_{\chi}$. As a reference point we have included an
estimate of the integrated exposure time after one calendar year of
running for some of the projected liquid xenon experiments in
Table~\ref{experiments}. These estimates assume 200 days of
data-taking per calendar year with 80\% of the nominal masses in
Table~\ref{experiments} being used as the fiducial mass.

%
%---------------- Pair 212 Table --------------------
\begin{table}[ht]
\begin{center}
\begin{tabular}{|l||cccccc|ccc|c|} \hline
 & \multicolumn{6}{c}{Mass Parameters (GeV)} &   \multicolumn{3}{c}{LSP Wave
 Function} & \\
 & $m_{N_1}$ & $m_{N_2}$ & $m_{C_1}$ & $m_{\tilde{\tau}_1}$
 & $m_h$ & $\mu$  & $\wtd{B}$\% & $\wtd{W}$\%
 & $\wtd{H}$\% & $\Ochi$ \\
\hline
Point~A & 237 & 240 & 239 & 261 & 117.4 & 991 & 78\% & 21\% & 1\% & 0.0054 \\
Point~B & 260 & 749 & 260 & 450 & 117.4 & 949 & 0\% & 99\% & 1\% & 0.0026  \\
\hline
%
%$m_{N_1}$ (GeV) & 237 & 260  \\
%%
%$m_{N_2}$ (GeV) & 240 & 749 \\
%%
%$m_{C_1}$ (GeV) & 239 & 260 \\
%%
%$m_{\tilde{\tau}_1}$ (GeV) & 261 & 450 \\
%%
%$m_h$ (GeV) & 117.4 & 117.4 \\
%%
%$\mu$ (GeV) & 991 & 949 \\
%%
%$\tan\beta$ & 37.3 & 49.5 \\ \hline
%%
%$\wtd{B}$\% & 78\% & 0\% \\
%%
%$\wtd{W}$\% & 21\% & 99\% \\
%%
%$\wtd{H}$\% & 1\% & 1\% \\
%%
%$\Ochi$ & 0.0054 & 0.0026 \\  \hline
%
\end{tabular}
\end{center}
{\caption{\label{pair212}\footnotesize {\bf Pair 212 of the 276
degenerate pairs}. Some key parameters for the two models making up
degenerate pair \#212. This particular case is an example of a
``squeezer'' degeneracy, in the language
of~\cite{ArkaniHamed:2005px}.}}
\end{table}
%------------------------- END OF THE TABLE ---------------------

Generally speaking when two models are visible they are easily
distinguished, at least under the idealized assumption of perfect
theoretical control over the input nuclear matrix elements.
Consider, for example, the pair of points described in
Table~\ref{pair212}. Both models are consistent with all
experimental constraints, including those of~(\ref{constraints}).
This model is typical of the set from Arkani-Hamed et al. In fact,
it is of the sort that were dubbed ``squeezers:'' a case in which
the mass difference between gauginos of the electroweak sector is
small for one of two models, making the decay products from gaugino
cascade decays too soft to be detected. The lepton-based signatures
are accidently similar because of the compensating change in the
stau masses. Both points predict a physically acceptable, if low,
thermal relic abundance of neutralinos. The low values are due to
the large wino-content of the LSP, and imply that detection rates
should be scaled downward by factors of~0.22 for Point~A and 0.1~for
Point~B. Consequently, despite relatively large spin-independent
interaction cross sections of $\sigma_{\chi\,p}^A =
3.61\times10^{-45}$ cm$^2$ and $\sigma_{\chi\,p}^B =
4.56\times10^{-45}$ cm$^2$, neither would produce any interactions
in current direct search experiments. We estimate that after one
ton-year of exposure, a liquid xenon based experiment would collect
$N_A = 1310$ recoil events for Point~A and $N_B = 1517$ recoil
events for Point~B if we do not rescale their local relic density
$n_{\chi}$.\footnote{Note that ``one ton-year of exposure'' is not
necessarily the same thing as ``after one year of data-taking at
XENON1T.'' If we assume data is taken roughly 200 days per year in a
fiducial volume of 80\% of the nominal volume, then it will take
roughly 27 months for a one-ton xenon-based detector to accumulate
this much exposure.} With rescaling these become $N_A = 282$ and
$N_B = 157$. With these latter numbers, and assuming zero background
contribution ($f=0$), these two signals differ by $6.4\sigma$. Even
taking $f=0.5$ they still differ by $5.3\sigma$.

The above case is typical; in a world without theoretical
uncertainties the limiting factor in distinguishing between these
degenerate pairs is the requirement of 100~events total for each
model in the pair. For example, with no rescaling of the local relic
density and setting $f=0$ in~(\ref{oursigma}) there are 231~pairs
for which both models in the pair give 100 events in one ton-year of
xenon exposure, versus 217~such pairs in germanium.\footnote{The
numbers are comparable in size, despite the difference in target
nuclei, because of the energy integration ranges~(\ref{rates}) we
have chosen.} If we require these signals to differ by two~sigma the
numbers become~190 and~176, respectively. Requiring five~sigma
significance only reduces these totals to~154 and~147. Now requiring
five~sigma significance {\em and} smearing the counts by a factor of
$f=0.5$ only reduces the totals to~144 and~134. In fact, of these
154~pairs that can be distinguished in one ton-year of liquid xenon
exposure (with $f=0$), the average separation significance is
$21.8\sigma$! The equivalent number for the 147~pairs
distinguishable in an equivalent exposure of germanium is
$24.4\sigma$. As soon as {\em both} of the models surpass our
threshold for a detectable ``signal'' they are almost always
immediately distinguishable. As we will see below, this high degree
of separability is an artifact of the assumption of perfect
theoretical precision on the calculation of interaction
cross-sections.

%---------------- Direct Detection Table I --------------------
\begin{table}[t]
\centering
\begin{tabular}{llcclcc|lcclcc}
 & \multicolumn{12}{c}{With Density Rescaling} \\ \hline \hline
 & \multicolumn{6}{c}{Require 100 Events} &   \multicolumn{6}{c}{Require 10
 Events} \\
 & & \multicolumn{2}{c}{Xenon} & & \multicolumn{2}{c}{Germanium} &
 &  \multicolumn{2}{c}{Xenon} & & \multicolumn{2}{c}{Germanium} \\
 \hline
 & & $f=0$ & $f=0.5$ & & $f=0$ & $f=0.5$ &  & $f=0$ & $f=0.5$ & & $f=0$ & $f=0.5$
 \\
 & $3\sigma$ & 8 & 8 & & 8 & 7 &  $3\sigma$ & 24 & 22 & & 24 & 21 \\
\raisebox{1.5ex}{0.1 ton-yr} & $5\sigma$ & 6 & 4 & & 5 & 3 &
$5\sigma$ & 14 & 9 & & 14 & 8 \\ \hline
 & $3\sigma$ & 79 & 71 & & 69 & 58 &  $3\sigma$ & 164 & 148 & & 157 & 136 \\
\raisebox{1.5ex}{1 ton-yr} & $5\sigma$ & 52 & 43 & & 48 & 37 &
$5\sigma$ & 112 & 81 & & 105 & 73 \\ \hline
 & $3\sigma$ & 199 & 182 & & 187 & 178 &  $3\sigma$ & 217 & 199 & & 212 & 201 \\
\raisebox{1.5ex}{5 ton-yr} & $5\sigma$ & 170 & 159 & & 162 & 151 &
$5\sigma$ & 187 & 175 & & 183 & 172 \\ \hline
\end{tabular}
%\end{center}
%
{\caption{\label{direct1}\footnotesize {\bf Direct detection summary
table}. We give the number of pairs distinguishable after a given
accumulated exposure in xenon or germanium, based on the integration
ranges specified in the text. All numbers in this table were
computed with interaction rates scaled by the quantity $r_{\chi} =
{\rm Min}(1,\, \Ochi/0.025)$.}}
\end{table}
%------------------------- END OF THE TABLE ---------------------

We summarize the numerical results of this section in
Tables~\ref{direct1} and~\ref{direct2}. We provide the number of
model pairs that can be distinguished at either the $3\sigma$ or
$5\sigma$ level in three different exposure times for xenon and
germanium. We continue to assume that theoretical uncertainties are
under control. In order to provide the reader with some context on
the additional assumptions we have made, we provide the data using a
threshold of 100~events and 10~events as well as calculating
distinguishability with a conservative error ($f=0.5$) or assuming
no background ($f=0$).  Finally, we note that of the 77~model pairs
we originally designated as ``physical'' in
Section~\ref{sec:constraints}, 57~of them will be distinguished at
the $5\sigma$ level (requiring a 100~event threshold) after one
ton-year of exposure in xenon, assuming $f=0$ and without rescaling
the local density of neutralinos. With rescaling only 3~of these
``physical'' models will be distinguished in one ton-year. Extending
the exposure time to five ton-years increases these numbers to
65~out of~77 (no rescaling) and 31~out of~77 (with rescaling).

%---------------- Direct Detection Table II --------------------
\begin{table}[h]
\centering
\begin{tabular}{llcclcc|lcclcc}
 & \multicolumn{12}{c}{Without Density Rescaling} \\ \hline \hline
 & \multicolumn{6}{c}{Require 100 Events} &   \multicolumn{6}{c}{Require 10
 Events} \\
 & & \multicolumn{2}{c}{Xenon} & & \multicolumn{2}{c}{Germanium} &
 &  \multicolumn{2}{c}{Xenon} & & \multicolumn{2}{c}{Germanium} \\
 \hline
 & & $f=0$ & $f=0.5$ & & $f=0$ & $f=0.5$ &  & $f=0$ & $f=0.5$ & & $f=0$ & $f=0.5$
 \\
 & $3\sigma$ & 51 & 41 & & 49 & 41 &  $3\sigma$ & 116 & 102 & & 111 & 98 \\
\raisebox{1.5ex}{0.1 ton-yr} & $5\sigma$ & 36 & 29 & & 32 & 25 &
$5\sigma$ & 81 & 69 & & 77 & 63 \\ \hline
 & $3\sigma$ & 177 & 168 & & 168 & 161 &  $3\sigma$ & 210 & 200 & & 208 & 198 \\
\raisebox{1.5ex}{1 ton-yr} & $5\sigma$ & 154 & 144 & & 147 & 134 &
$5\sigma$ & 181 & 165 & & 175 & 155 \\ \hline
 & $3\sigma$ & 242 & 235 & & 240 & 234 &  $3\sigma$ & 242 & 235 & & 240 & 234 \\
\raisebox{1.5ex}{5 ton-yr} & $5\sigma$ & 224 & 216 & & 217 & 211 &
$5\sigma$ & 224 & 216 & & 217 & 211 \\ \hline
\end{tabular}
%\end{center}
%
{\caption{\label{direct2}\footnotesize {\bf Direct detection summary
table}. Same as Table~\ref{direct1}, only without scaling
interaction rates. }}
\end{table}
%------------------------- END OF THE TABLE ---------------------

Tables~\ref{direct1} and~\ref{direct2} illustrate the incredible
power of the next-generation of dark matter direct detection
experiments. The improvements to (already good) background rejection
mechanisms will make these experiments very accurate counters -- but
without equivalent improvements on the issue of theoretical inputs
the utility of these measurements will be severely degraded. This
point was raised by Ellis et al.~\cite{Ellis:2008hf}, who studied
the variation in the prediction for $\sigma^{\mathrm{SI}}_{\chi p}$
arising from differing input values of the $\pi\,N$ $\Sigma$-term.
We computed the value of $\sigma^{\mathrm{SI}}_{\chi p}$ for the
378~models of this study using the range of input parameters
considered in Ref.~\cite{Ellis:2008hf}, arriving at a typical
variation of approximately 50\% for the interaction cross-section.
This is similar to the observed variation of Ellis et al., as well
as others who have undertaken similar
investigations~\cite{Barger:2007nv}. Let us consider an additional
theoretical error proportional to the calculated cross-section
\begin{equation} \(\delta \sigma^{\mathrm{SI}}_{\chi} \)_{\rm theor}
= \epsilon \times \sigma^{\mathrm{SI}}_{\chi} \label{epsilon}
\end{equation}
which is to be added in quadrature to the statistical errors
considered previously. In Table~\ref{theoryunc} we show the
degradation in the number of distinguishable pairs that arises from
considering $\epsilon \neq 0$. The entries in Table~\ref{theoryunc}
should be compared to those of the right panel in
Figure~\ref{direct1}; we require at least 10~signal events in one
(or five) ton-years of xenon exposure. The interaction rates in this
table were rescaled for cases of low predicted thermal relic
density. Clearly if theoretical uncertainties stay at their present
level (with roughly 50\% uncertainty in the cross-section
predictions) then it will be impossible to distinguish models with
direct detection experiments -- even after five ton-years of
exposure and requiring only $3\sigma$ separation. If the uncertainty
in the $\pi\,N$ $\Sigma$-term can be reduced so as to generate only
a 10\% theoretical uncertainty in $\sigma^\mathrm{SI}_{\chi p}$ the
ability to distinguish models will still be significantly reduced,
but some hope for separating models will remain. For this reason we
strongly echo the call made by Ellis et al. for further experimental
work aimed towards reducing these uncertainties.

%---------------- Theoretical Error Table --------------------
\begin{table}[h]
\centering
\begin{tabular}{llcccc}
\hline \hline
 & \multicolumn{5}{c}{Require 10 Events, Xenon} \\
\hline
 & & $\epsilon=0$ & $\epsilon=0.1$ & $\epsilon=0.25$ & $\epsilon=0.5$ \\
 & $3\sigma$ & 164 & 118 & 13 & 0 \\
\raisebox{1.5ex}{1 ton-yr} & $5\sigma$ & 112 & 46 & 0 & 0 \\
\hline
 & $3\sigma$ & 217 & 149 & 25 & 0 \\
\raisebox{1.5ex}{5 ton-yr} & $5\sigma$ & 187 & 77 & 0 & 0 \\
\hline
\end{tabular}
%\end{center}
%
{\caption{\label{theoryunc}\footnotesize {\bf Effect of theoretical
uncertainties associated with nuclear matrix elements}. We give the
number of pairs distinguishable after a given accumulated exposure
in xenon when an additional theoretical uncertainty of the form
of~(\ref{epsilon}) is included. We take the experimental error to be
purely statistical ($f=0$) and require 10~signal events. All numbers
in this table were computed with interaction rates scaled by the
quantity $r_{\chi} = {\rm Min}(1,\, \Ochi/0.025)$. }}
\end{table}
%------------------------- END OF THE TABLE ---------------------

\section{Gamma Ray Experiments}
\label{sec:gamma}

Direct detection experiments measure the presence of relic
neutralinos at the location of the earth. Indirect detection of
relic neutralinos involves looking for the products of neutralino
pair annihilation processes far from the earth's location. Given a
supersymmetric model, the rate for annihilation into various final
states can be computed as a function of parameters such as the
neutralino mass $m_{\chi}$. Of these possible final states, photons
in the gamma ray energy regime are unique in that they are
uninfluenced by galactic magnetic fields and travel largely
unimpeded from their source. This allows gamma ray observatories to
concentrate on areas of the sky likely to have a high relic
neutralino density -- such as the center of our galaxy. This is
important since the rate for annihilation processes scales like the
square of the local density $\rho_{\chi}$. Photons can be produced
as part of decay chains (for example, from the decays of neutral
pions in hadronic decays) or directly through loop-induced diagrams.
The former contribute to a continuous general spectrum of photons
and are thus more difficult to distinguish from astrophysical
backgrounds. But direct production of two photons -- or production
of a single photon in association with a Z-boson -- through
loop-diagrams offers the possibility of a monochromatic spectrum
that can more easily be distinguished from the
background~\cite{Bertone:2004pz}. The trade-off is a reduction in
rate relative to the continuous photon rate.

The rate of gamma ray photons observed at the earth's location is
sensitive to the radial dependence of the relic neutralino density
profile assumed. The majority of the profiles considered in the
literature can be described by a common parameterization
\beq \rho_{\chi}(r) = (\rho_{\chi})_0
\frac{\(r/r_0\)^{-\gamma}}{\[1+\(r/a\)^{\alpha}\]^{\frac{\beta-\gamma}{\alpha}}}
\[1 + \(\frac{r_0}{a}\)^{\alpha}\]^{\frac{\beta-\gamma}{\alpha}} \,
,
\label{density} \eeq
where $r_0$ is roughly the distance between the earth and the
galactic center and the normalization constant $(\rho_{\chi})_0$ is
typically taken to be 0.3~GeV/cm$^3$. It is this parameter which is
rescaled for those models with $\Ochi < 0.025$. The function
in~(\ref{density}) is integrated along the line of sight between the
earth and the observed object -- for us this will be the galactic
center. All of the physics of the halo model chosen is therefore
sequestered to a single overall scaling parameter $J(\psi)$
\beq J(\psi) = \frac{1}{8.5\,{\rm kpc}}\(\frac{1}{0.3 \GeV\,{\rm
cm}^3}\)^2 \int_{\rm line\, of\, sight} ds(\psi) \, \rho^2_{\chi}
(r) \, \label{Jpsi} \eeq
with $s(\psi)$ being a parameter running along the ray to the
observed object, making an angle $\psi$ relative to the galactic
center. For observations of the galactic center $\psi=0$. If the
quantity $\Delta \Omega$ represents the finite angular resolution of
a given detector, then the quantity~(\ref{Jpsi}) is integrated over
a spherical region of solid angle $\Delta \Omega$ centered on
$\psi=0$. The average value of $J(\psi)$ over this region
\beq \lang J(\psi) \rang_{\Delta \Omega} = \frac{1}{\Delta
\Omega}\int_{\Delta \Omega} d\Omega' \, J(\psi') \label{Jave} \eeq
is often denoted $\oline{J}(\Delta \Omega)$ and is handy for
comparing rates computed with different halo model assumptions.

Different choices for the parameter set
$(a,\,r_0,\,\alpha,\,\beta,\,\gamma)$ common in the literature can
change the resulting value of $\oline{J}(\Delta \Omega)$ (and hence
the predicted gamma ray flux) by several orders of magnitude. This
is by far the greatest source of uncertainty in making definitive
statements about the ability of gamma ray experiments to distinguish
between competing theories. Indeed, values of $\oline{J}(\Delta
\Omega) \gappeq {\rm few} \times 10^{4}$ will be required if any of
the current or proposed experiments are to observe a signal at all.
In this paper we will therefore primarily consider the model of
Navarro, Frank and White (NFW)~\cite{Navarro:1995iw} as well as a
modified NFW profile that takes into account the effect of baryons
(a so-called ``adiabatic compression'' (AC)
model)~\cite{Blumenthal:1985qy,Mambrini:2005vk}.
On occasion we will also refer to a more singular profile by Moore
et al.~\cite{Moore:1999nt}, also with the effects of adiabatic
compression, which is more favorable for the prospects of indirect
dark matter detection at gamma ray experiments.
The parameters associated with these three models, as well as the
computed values of $\oline{J}(\Delta \Omega)$ for the galactic
center are given in Table~\ref{halomodel}. These values can be used
to scale our subsequent results for comparison to other profile
models.

%---------------- Halo Model Table --------------------
\begin{table}[th]
\begin{center}
\begin{tabular}{|c||c|c|c|c|c||c|} \hline
Model & $r_0$ (kpc) & $a$ (kpc) & $\alpha$ & $\beta$ & $\gamma$
& $\oline{J}(10^{-5}\,{\rm sr})$\\
\hline
NFW & 8.0 & 20.0 & 1 & 3 & 1 & $1.2644 \times 10^4$\\ \hline
NFW + AC & 8.0 & 20.0 & 0.8 & 2.7 & 1.45 & $1.0237\times 10^6$ \\
\hline
%
%Moore & 8.0 & 28.0 & 1.5 & 3 & 1.5 & 9.68\times10^{6}\\
%
Moore + AC & 8.0 & 28.0 & 0.8 & 2.7 & 1.65 & $3.0896\times10^8$ \\
\hline
\end{tabular}
\end{center}
{\caption{\label{halomodel}\footnotesize {\bf Halo model
parameters}. The parameters which define the three halo models we
will consider in this work and the resulting value of the parameter
$\oline{J}(\Delta \Omega)$ for $\Delta\Omega = 10^{-5}$ sr. We will
primarily consider the NFW halo profile with adiabatic compression
in what follows.}}
\end{table}
%------------------------- END OF THE TABLE ---------------------

Before proceeding we emphasize that we will treat the three input
values for $\oline{J}(\Delta \Omega)$ in Table~\ref{halomodel} as
discrete possibilities for this crucial input parameter. It is
unclear whether this is a reasonable assumption, though it is
commonly made in the literature. As we will argue at the end of this
section, consistency between any post-LHC supersymmetric model and
any future signal of dark matter annihilation at the galactic center
will likely single out only one of these profile models given their
wide disparity in $\oline{J}(\Delta \Omega)$ values. But if these
values merely represent signposts along a {\em continuum} of
possible $\oline{J}(\Delta \Omega)$ it is unclear how indirect
detection of dark matter via annihilation into photons will ever be
able to distinguish between two candidate supersymmetric models --
at least in the absence of other, orthogonal data from other
cosmological observations. The situation would then be analogous to
trying to separate two theories on the basis of the number of
trilepton events at the LHC, but with a luminosity that is uncertain
by as much as four orders of magnitude! Clearly, then, maximizing
the value of future experimental observations will require much
better knowledge of the uncertainties associated with numbers such
as those in Table~\ref{halomodel}. We will continue this section
under the original assumption that only one of these models is
correct and that the prediction for $\oline{J}(\Delta \Omega)$
associated with that model is exact, returning to the issue of
theoretical uncertainty at the end of the section.

The differential flux of photons from neutralino annihilation at the
earth's location is then given by
\beq \frac{d\Phi_{\gamma}(E_{\gamma})}{dE_{\gamma}} = \sum_{i}
\frac{\lang \sigma_i v \rang}{4\pi m_{\chi}^2}
\frac{dN^i_{\gamma}}{dE_{\gamma}} \int_{\rm line\, of\, sight}
ds(\psi) \, \rho^2_{\chi} (r)\, , \label{flux1} \eeq
where the summation is over all possible final states, $\sigma_i$ is
the annihilation cross-section into that final state, and
$dN^i_{\gamma}/dE_{\gamma}$ is the differential spectrum of photons
produced in the decay channel labeled by the index $i$. Replacing
the halo integral in~(\ref{flux1}) by~(\ref{Jave}) gives
\beq \frac{d\Phi_{\gamma}(E_{\gamma})}{dE_{\gamma}} = 0.94 \times
10^{-13} \sum_i \frac{dN^i_{\gamma}}{dE_{\gamma}} \(\frac{\lang
\sigma_i v \rang}{10^{-29}\, {\rm cm}^3\,{\rm
s}^{-1}}\)\(\frac{100\GeV}{m_{\chi}}\)^2 \oline{J}(\Delta \Omega)
\Delta\Omega  \label{flux2} \eeq
in units of photons/cm$^2$/s/GeV. The final step is to perform an
integration over the energy range relevant to the particular
experiment. The continuous gamma ray spectrum from neutralino
annihilation has a sharp cut off at the mass of the neutralino
$m_{\chi}$. We therefore integrate~(\ref{flux2}) from some
sensitivity threshold $E_{\rm min}$ to $E_{\rm max}$, where $E_{\rm
max}$ is the smaller of the mass of the neutralino or the upper
limit of the experiment's energy sensitivity. Note that for
monochromatic lines produced by processes $\chi_0\, \chi_0 \to
\gamma\,\gamma$ and $\chi_0\, \chi_0 \to \gamma\,Z$ no such
integration is necessary and $dN^i_{\gamma}/dE_{\gamma}$
in~(\ref{flux2}) is replaced by $N_{\gamma} = 2$ and $N_{\gamma} =
1$ for these two cases, respectively.

%%%%%%%%%%%%%%%%%%%%%%%%%%%%%%%%%%%%%%%%%%

In this section we will consider two broad classes of gamma ray
observatories which are operational now or will become so in the
near future: space-based satellites such as the GLAST
LAT~\cite{Gehrels:1999ri} and ground based atmospheric Cherenkov
telescopes (ACTs) such as CANGAROO~\cite{Cangaroo},
HESS~\cite{Hess}, MAGIC~\cite{Magic} and VERITAS~\cite{Veritas}. The
two classes have different strengths and weaknesses, making them
suited for different signals of dark matter annihilation.
Space-based experiments can observe nearly continuously, while
ground based experiments can collect data only on dark, cloudless
nights. On the other hand these ACTs cover a much larger effective
area $A_{\rm eff}$ of the sky than the GLAST LAT. The GLAST
experiment will be sensitive to photon energies from
$\order(10\,{\rm MeV})$ to about 300~GeV -- a photon energy range
much lower than that of the ACTs whose threshold energies are in the
range 50-100 GeV, if not higher. The lower energy region is more
suited for measurements of the continuous gamma ray flux arising
from neutralino annihilations. The masses of the neutralino LSP in
our 378~models fall in the range
\beq 98 \GeV \leq m_{\chi} \leq 557 \GeV \label{massrange} \eeq
with over 85\% having a mass less than 300~GeV. We will therefore
discuss the continuous gamma ray spectrum only in the context of the
GLAST experiment. However, the two monochromatic fluxes associated
with di-photon and photon/Z-boson final states have associated
photon energies
\bea E_{\gamma \gamma} &=&  m_{\chi} \nonumber \\
E_{\gamma Z} &=& m_{\chi} - M_Z^2/4m_{\chi}\, , \label{lines} \eea
respectively. For these signals we will only consider ACT
experiments; GLAST is sensitive to many of the relevant energy
values but the flux is too small to give an appreciable rate in all
but the Moore (AC) halo profile. In Table~\ref{gamexp} we list the
relevant experimental parameters for GLAST and a generic ACT based
roughly on the characteristics of the HESS and VERITAS experiments.
We will use these two experimental configurations in our analysis.

%%%%%%%%%%%%%%%%%%%%%%%%%%%%%%%%%%%%%%%%%%

%---------------- Gamma Ray Experiments Table --------------------
\begin{table}[th]
\begin{center}
\begin{tabular}{|l||ccccc|} \hline
 & $E_{\rm min}$ & $E_{\rm max}$ & $\sigma_E/E$ & $A_{\rm eff}$ & $\Delta
 \Omega$ \\ \hline \hline
GLAST & 50~MeV & 300~GeV & 10\% & $1\times10^4$ cm$^2$ &
$1\times10^{-5}\,{\rm sr}$ \\
ACT & 100~GeV & 10~TeV & 15\% & $3\times10^8$ cm$^2$ &
$1\times10^{-5}\,{\rm sr}$ \\ \hline
%
%  & \parbox{3cm}{\begin{center}GLAST\end{center}} &
%  \parbox{3cm}{\begin{center}ACT\end{center}} \\ \hline \hline
%%
%$E_{\rm min}$ & 50 MeV &  100 GeV \\
%% GLAST: Zaharijas & Hooper (2006) say 100 MeV
%% MAGIC: Sanchez-Conde talk (DM2008) says 60-100 GeV
%% VERITAS: website says 50-100 GeV
% MAGIC: Bastieri talk (Rome2007) says 50 GeV
%
%$E_{\rm max}$ & 300 GeV & NA \\
%
%$\sigma_E/E$ & 10\% & 15\% \\
% VERITAS: astro-ph/0709.3654 10-20%
% MAGIC: Bastieri talk (Rome2007) says 20-30%
%
%$A_{\rm eff}$ & $1\times10^4$ cm$^2$ & $3\times10^8$ cm$^2$ \\
% HESS/VERITAS: Bergstrom et al. (1997) estimates 7x10^8 cm^2
% VERITAS: astro-ph/0709.3654 >3x10^4 m^2
%$\Delta \Omega$ & $1\times10^{-5}\,{\rm sr}$ & $1\times10^{-5}\,{\rm
%sr}$ \\ \hline
% HESS: Moulin talk (SUSY07) 2x10^-5 sr
%
% HESS: Zaharijas & Hooper (2006) say 80 hours exposure time
% HESS: Moulin talk (SUSY07) 1000hrs/year
%
\end{tabular}
\end{center}
{\caption{\label{gamexp}\footnotesize {\bf Gamma Ray Experiments}.
For the analysis presented in this section we consider two gamma ray
observatories: the GLAST satellite-based experiment and a generic
atmospheric Cherenkov telescope. }}
\end{table}
%------------------------- END OF THE TABLE ---------------------

An important distinction between the gamma ray experiments and the
nuclear recoil experiments of Section~\ref{sec:direct} is in the
nature of backgrounds. We argued in the previous section that direct
detection experiments expect to achieve a high degree of background
rejection in future incarnations. For our analysis of
distinguishability we assumed a small (but non-vanishing) rate for
misidentifying nuclear recoils. The rates we have in mind for these
experiments were in the range of 1~-~10~events per experiment per
year. This motivated the requirement of at least 10 to 100~nuclear
recoil events observed for a given model before a definitive
statement is made. There is no similarly successful method for
discriminating between photons resulting from neutralino
annihilation and those that arise from other astrophysical processes
that generate gamma rays of the same energy range. As a result,
indirect detection of relic neutralinos implies observing an excess
of high energy photons above some background rate that can be quite
substantial. Accurate and believable modeling of this background
photon rate is therefore crucial for making measurements that can be
used to distinguish between two models in a degenerate pair.

Photons in the gamma ray energy regime are produced from a number of
sources. These include high energy cosmic ray proton collisions with
helium gas which produces $\pi^0$'s, interactions of electrons with
galactic radiation via inverse Compton scattering and bremsstrahlung
processes from accelerated charges~\cite{Strong:2007nh}. For
energies in the broad range $100\,{\rm MeV} \lappeq E_{\gamma}
\lappeq 1\,{\rm TeV}$ relevant for neutralino annihilation, the sum
of these processes produces a differential spectrum typically
modeled by a power law of the form~\cite{Bergstrom:1997fj}
\beq \frac{d^2\Phi^{\rm bkg}_{\gamma}}{d\Omega dE_{\gamma}} =
\(\frac{d\Phi^{\rm bkg}_{\gamma}}{d\Omega
dE_{\gamma}}\)_0\,\(\frac{E_{\gamma}}{1~\GeV}\)^{-2.7} \,
\label{bkgrnd} \eeq
where $(d\Phi^{\rm bkg}_{\gamma}/d\Omega\,dE_{\gamma})_0$ has units
of photons/cm$^2$/s/sr/GeV and serves as a normalization for the
spectrum. A reasonable value for this parameter would be
approximately
\beq \(\frac{d^2\Phi^{\rm bkg}_{\gamma}}{d\Omega dE_{\gamma}}\)_0
\simeq 9 \times 10^{-5} \,{\rm photons/cm^{2}/s/sr/GeV}
\label{lownorm} \eeq
in the direction of the galactic center. Integrating~(\ref{bkgrnd})
with normalization~(\ref{lownorm}) over a typical range relevant to
GLAST of $1\GeV \leq E_{\gamma} \leq 200\GeV$, and using $\Delta
\Omega = 10^{-5}$, one obtains approximately 100 background photons
per m$^2$-year of exposure. This suggests that any model capable of
producing a flux from neutralino annihilations of the order of
$\Phi_{\gamma} \simeq 10^{-10}\,{\rm photons/cm^{2}/s}$ at GLAST
(when integrated from a threshold energy of 1~GeV) should be
detectable above background. Such estimates are often quoted in the
literature~\cite{Morselli:2002nw,Hooper:2003ka}.

%As their name suggests, atmospheric Cherenkov telescopes detect
%cosmic gamma rays only indirectly, by observing a shower of
%secondary particles in the earth's atmosphere. Therefore showers
%initiated by cosmic ray electrons are largely indistinguishable from
%those with a photon primary, except through inference via the
%direction of arrival. This is an additional source of backgrounds
%for these experiments which we will not consider in this work.

However, there are reasons to be concerned about how well this
background rate is understood~\cite{Strong:2007nh,Zaharijas:2006qb}.
Observations from the satellite-based EGRET
experiment~\cite{MayerHasselwander:1998hg} indicated a
higher-than-expected photon flux in the energy range of 100~MeV to
approximately 10~GeV in the direction of the galactic center.
Furthermore, earth-based ACT experiments
HESS~\cite{Aharonian:2004wa}, VERITAS~\cite{Kosack:2004ri},
CANGAROO~\cite{Tsuchiya:2004wv} and MAGIC~\cite{Albert:2005kh} also
observed an anomalously large gamma-ray source in the direction of
the center of the galaxy for higher energy photons ($200 \GeV
\lappeq E_{\gamma} \lappeq 10 \TeV$). It is likely that both the
EGRET data and the ACT data represent newly identified point-like
sources near the galactic
center~\cite{Hooper:2002ru,Aharonian:2004jr}. Given the fine angular
resolution available with GLAST it should be possible to subtract
these point sources from the more diffuse signal expected from
neutralino annihilation. This issue was the subject of a recent
study by Dodelson, Hooper and Serpico~\cite{Dodelson:2007gd} in
which the separation of dark matter signals from background sources
was performed using both spectral information as well as angular
information. We will loosely follow the example
of~\cite{Dodelson:2007gd} for our treatment of backgrounds in light
of the EGRET/ACT data.

We therefore consider two different background estimations. The
first is simply the one represented by the power law
in~(\ref{bkgrnd}) with normalization~(\ref{lownorm}). This is the
standard case studied in the literature and we will refer to this
case as the ``low'' background. For the second background estimation
we will use~(\ref{bkgrnd}) and~(\ref{lownorm}) as a baseline but
then take into account the EGRET data by adding to it an additional
contribution modeled by~\cite{Dodelson:2007gd}
\beq \frac{d\Phi^{\rm EG}_{\gamma}}{dE_{\gamma}} = 2.2 \times
10^{-7}\,\times{\rm
exp}\(-\frac{E_{\gamma}}{30\GeV}\)\times\(\frac{E_{\gamma}}{1\GeV}\)^{-2.2}
\,\, {\rm photons/cm^{2}/s/GeV} \, . \label{EGRET} \eeq
For the high energy ACT sources we add an additional contribution
modeled by~\cite{Dodelson:2007gd}
\beq \frac{d\Phi^{\rm ACT}_{\gamma}}{dE_{\gamma}} = 1.0 \times
10^{-8}\,\times\(\frac{E_{\gamma}}{1\GeV}\)^{-2.25} \,\, {\rm
photons/cm^{2}/s/GeV} \, , \label{ACT} \eeq
for energies $E_{\gamma} \geq 200\GeV$. In practice we will only
integrate the GLAST signal up to 200~GeV, so this second
contribution in~(\ref{ACT}) will be relevant only as background to
monochromatic signals in the generic ACT~experiment of
Table~\ref{gamexp}. This set of assumptions we will refer to as the
``high'' background. We note again that these additional
contributions~(\ref{EGRET}) and~(\ref{ACT}) presumably represent
point sources which may be reliably removed from the data given the
angular resolution of GLAST. For our simple study we will not bin
the data in solid angle, and perform only a crude binning in energy.
We therefore consider treating~(\ref{EGRET}) and~(\ref{ACT}) as
additional diffuse contributions to be a reasonable and conservative
assumption.

Let us begin with the continuous differential spectrum and the GLAST
experiment. Following~\cite{Dodelson:2007gd} we will attempt to take
into account the shape of this spectrum as part of our analysis. For
each model we use {\tt DarkSUSY} to compute the differential photon
flux from~(\ref{flux2}) in 1~GeV increments over the energy range
$1\GeV \leq E_{\gamma} \leq 200\GeV$. From this information we
create an interpolating function which is then integrated over six
energy bins: 1~-~10~GeV, 10~-~30~GeV, 30~-~60~GeV, 60~-~100~GeV,
100~-~150~GeV, and 150~-~200~GeV. For each model we also compute a
total integrated flux by integrating over the entire range $1\GeV
\leq E_{\gamma} \leq 200\GeV$. To say that two different signals are
detectable and distinguishable after a certain observation time we
require the following simultaneous conditions:
\begin{enumerate}
\item The total number of gamma ray photons $N_{\gamma}$ collected
by the experiment over the full energy range $1\GeV \leq E_{\gamma}
\leq 200\GeV$ must satisfy $N_{\gamma}>100$. We require this to be
true of {\em both} models in the model pair.
\item We require in addition that a significant excess of gamma ray
photons above background is observed in {\em multiple, adjacent}
energy bins. The premise behind this requirement is the desire to
have some spectral information on the component of the flux arising
from dark matter annihilation to better separate this source from
other astrophysical sources. Specifically, if $i =1,\dots,6$ labels
our six energy bins, then we demand  $N_i>m\sqrt{N_i^{\rm bkg}}$ for
at least three adjacent bins. Here $N_i$ is the number of photons
observed in that energy bin, $N_i^{\rm bkg}$ is the expected
background count (computed either with the ``low'' or the ``high''
background model), and $m$ is the significance level in units of
signal/$\sqrt{{\rm bkgrnd}}$. In what follows we will demand $m=2$.
\item If the above two conditions are satisfied by models $A$ and $B$
then we will say that the two potential signals are detectable. We
will further say that they are distinguishable if the condition
$|N_i^A-N_i^B|>n\sqrt{N_i^A+N_i^B+2N_i^{\rm bkg}}$ holds for at
least three adjacent bins, simultaneously. Results will be given for
significance levels $n=3$ and $n=5$.
%
%\item Finally, if the above are satisfied for different sets of 3 adjacent
%bins, the total number of model pairs that is distinguishable is
%found from the union of the possible sets of 3 adjacent bins (i.e.
%only count distinguishable pairs once if found in different sets of
%adjacent bins)
%
\end{enumerate}

%==================================================================
\begin{figure}[t]
%    \begin{center}
\centerline{
       \psfig{file=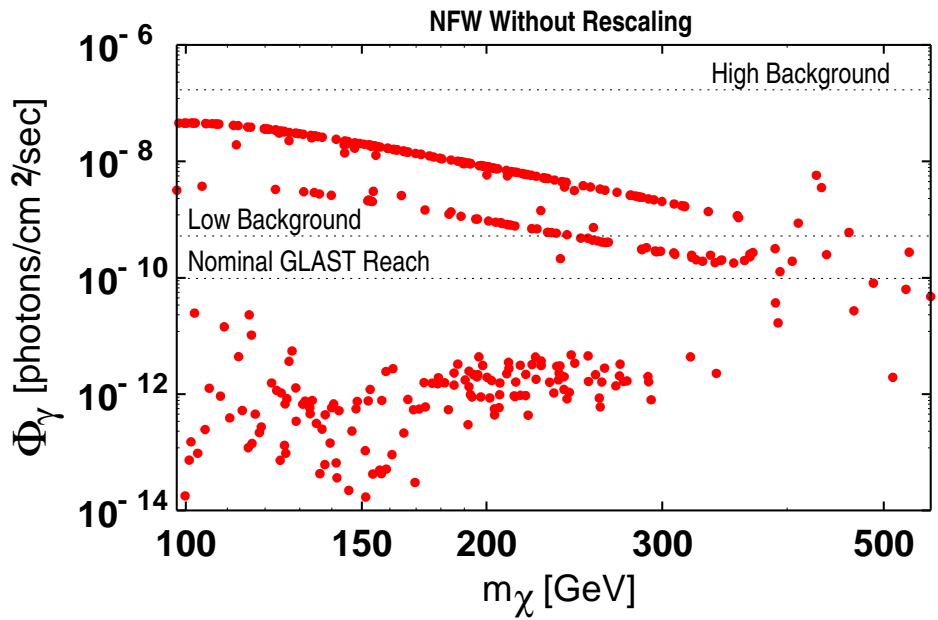,width=0.75\textwidth}
      }
          \caption{{\footnotesize {\bf Integrated photon flux for
          NFW profile and no density rescaling}. The differential
          photon flux is
          integrated over the energy range $1\GeV \leq E_{\gamma}
          \leq 200\GeV$ for the NFW profile. In this plot
          none of the rates have been rescaled in accordance
          with the predicted thermal relic density. The flux from background
          sources are indicated by the horizontal dotted lines for our
          ``low'' and ``high'' background models. We also give the nominal
          sensitivity threshold claimed by the GLAST collaboration.}}
        \label{fig:nfwflux}
%    \end{center}
\end{figure}
%======================================================================

\begin{figure}[t]
%    \begin{center}
\centerline{
       \psfig{file=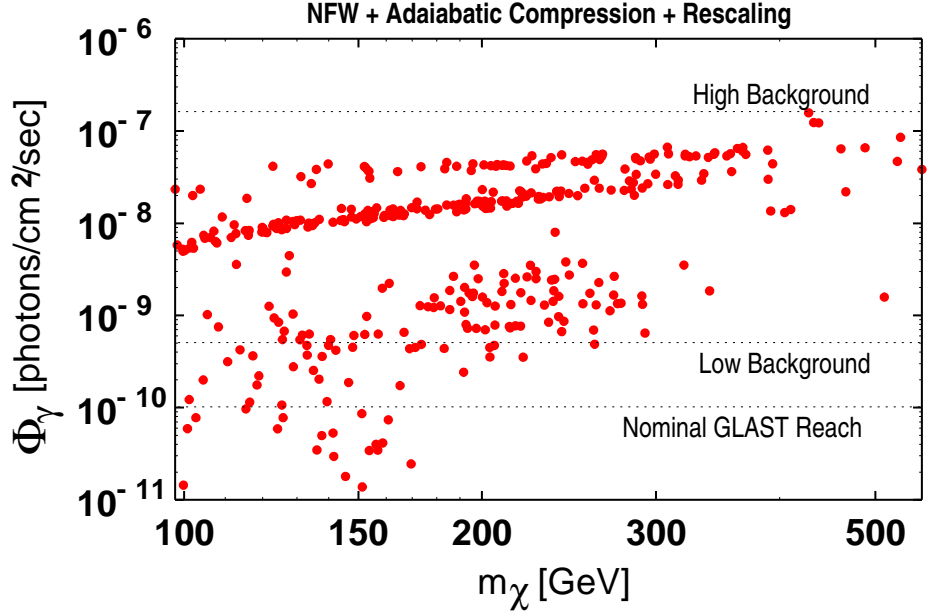,width=0.75\textwidth}
      }
          \caption{{\footnotesize {\bf Integrated photon flux for
          NFW profile + adiabatic compression}.
          Same as Figure~\ref{fig:nfwflux} but with the addition of
          adiabatic compression. In this plot we have rescaled the
          annihilation rate by the factor $r_{\chi}$.}}
        \label{fig:nfwacflux}
%    \end{center}
\end{figure}

The majority of studies in the literature focus on only the first of
the above items. Since the total (integrated) flux
$\Phi_{\gamma}(E_{\gamma} \geq 1\GeV)$ is such a commonly considered
variable we begin here. In Figure~\ref{fig:nfwflux} we plot the
integrated flux from $1\GeV \leq E_{\gamma} \leq 200\GeV$ arising
from dark matter annihilations as a function of the neutralino mass
$m_{\chi}$ in the NFW~halo profile. The typical integrated flux
values for our 378~models are on the order of~$10^{-11}$
photons/cm$^2$/sec if we rescale the local halo density by the
factor $r_{\chi}$, as is the default setting in {\tt DarkSUSY}.
Therefore if any significant flux is to be produced at all with the
NFW~profile then we must assume that the halo density
$\rho_{\chi}(r)$ is {\em not} rescaled. That was the assumption that
went into producing Figure~\ref{fig:nfwflux}.
Our two background estimations have been integrated over the same
energy range and are shown as the horizontal dotted lines. Note that
the ``low'' background rate corresponds well with the typically
quoted GLAST sensitivity limit of $\Phi_{\gamma} = 10^{-10}$
photons/cm$^2$/sec~\cite{Morselli:2002nw}. An alternative -- and
perhaps more reasonable -- set of assumptions is to add the prospect
of adiabatic compression but include the effects of halo density
rescaling. With these assumptions we get the distribution in
Figure~\ref{fig:nfwacflux}. Clearly more favorable assumptions could
be made; for example the profile of Moore et~al. with adiabatic
compression would boost these numbers by a factor of~roughly 30.
%
%We prefer to be more conservative while still making assumptions
%that allow for observations. We will therefore use the NFW~profile
%without rescaling and the NFW~+~adiabatic compression profile (with
%and without rescaling) in what follows, with an occasional reference
%to what might be possible with even more favorable halo assumptions.

%==================================================================
%======================================================================

\begin{figure}[t]
%    \begin{center}
\centerline{
       \psfig{file=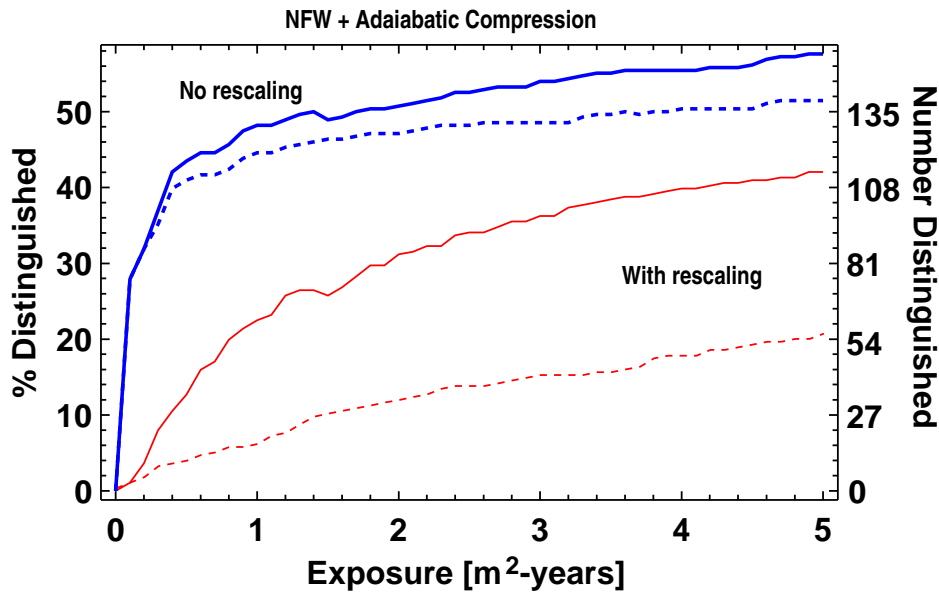,width=0.75\textwidth}
      }
          \caption{{\footnotesize {\bf Distinguishability analysis at
          GLAST}. The lower two curves rescale the local
halo density by the factor $r_{\chi}$ while the upper two curves
have no rescaling. The solid lines give the number of distinguished
pairs using the ``low'' background estimate, while dashed lines are
for the ``high'' background estimate.}}
        \label{fig:gamma}
%    \end{center}
\end{figure}

After converting the differential fluxes into actual photon counts
as described above, we ask how well the model pairs can be separated
as GLAST observation time is integrated. Our results are shown in
Figure~\ref{fig:gamma} as a function of integrated time-on-target in
units of m$^2$-years. Note that as in the case of direct detection
experiments one calendar year of the GLAST mission does not
necessarily produce 1~m$^2$-year of integrated observation since,
for example, the telescope will not be directed at the galactic
center continuously. The figure is computed for the NFW~halo profile
with adiabatic compression. The lower two curves rescale the local
halo density for those models where $\Ochi < 0.025$ while the upper
two curves have no such rescaling. Dashed curves count the number of
separable models using the low background, while solid curves use
the higher EGRET-normalized background. With the most conservative
assumptions -- rescaling the halo density for low $\Ochi$ models and
using the higher background estimate -- only 20\% of the model pairs
are distinguished even after 5~m$^2$-years. Assuming only the
``low'' background of~(\ref{bkgrnd}) and~(\ref{lownorm}) gives a
figure above 40\%. We feel this is a reasonable, if conservative,
estimate of the power of the GLAST experiment to distinguish between
degenerate pairs in the event that signals are visible at all above
background.
%

%==================================================================
%======================================================================

We summarize the results of the continuous gamma ray observables in
Table~\ref{gamma1}. Four different halo assumptions are listed: NFW
with rescaling the halo density, NFW~+~adiabatic compression (with
and without rescaling the halo density), and Moore et al. with halo
density rescaling. All results assume $m=2\sigma$, but we consider
both a high and low background assumption with either $3\sigma$ or
$5\sigma$ separation criterion.

% ---------------- CONTINUOUS GAMMA TABLE ----------------------
\begin{table}[h]
\begin{center}
\begin{tabular}{llcc|cc|cc|cc}
 & & \multicolumn{2}{c}{} & \multicolumn{2}{c}{NFW} &
 \multicolumn{2}{c}{NFW} & \multicolumn{2}{c}{Moore} \\
 & & \multicolumn{2}{c}{NFW} & \multicolumn{2}{c}{adiab. comp.} &
 \multicolumn{2}{c}{adiab. comp.} & \multicolumn{2}{c}{adiab. comp.} \\
 & & \multicolumn{2}{c}{not rescaled} & \multicolumn{2}{c}{rescaled} &
 \multicolumn{2}{c}{not rescaled} & \multicolumn{2}{c}{rescaled} \\
 \hline \hline
\multicolumn{2}{c}{Background:} & low & high & low & high & low &
high & low & high
 \\ \hline
& $3\sigma$ & 4 & 0 & 98 & 29  & 148 & 133 & 220 & 165 \\
\raisebox{1.5ex}{ 1 m$^2$ yr }
& $5\sigma$ & 0 & 0 & 60 & 14  & 131 & 122 & 196 & 151 \\
\hline
 & $3\sigma$ & 22 & 0 & 135 & 49  & 160 & 145 & 232 & 186 \\
\raisebox{1.5ex}{3 m$^2$ yr}
& $5\sigma$ & 9 & 0 & 95 & 33 & 144 & 134 & 185 & 173 \\
\hline
& $3\sigma$ & 35 & 0 & 147 & 62  & 167 & 147 & 207 & 194 \\
\raisebox{1.5ex}{5 m$^2$ yr}
& $5\sigma$ & 13 & 0 & 111 & 42  & 154 & 139 & 194 & 183 \\
\hline
\end{tabular}
\\
{\caption{\label{gamma1}\footnotesize {\bf Integrated gamma ray flux
summary table}. The number of pairs distinguishable at
the~$n=3\sigma$ and~$5\sigma$ level are listed for three different
integrated exposures at GLAST. The four halo models assumptions are
NFW without halo rescaling, NFW plus adiabatic compression (with and
without halo rescaling) and Moore et al. with halo rescaling. We
consider both the low (or standard) background estimate as well as
an EGRET-normalized higher background estimate. Theoretical
uncertainties associated with these halo models are neglected in
this table.}}
\end{center}
\end{table}

%%%%%%%%%%%%%%%%%%%%%%%%%%%%%%%%%%%%%%%%%%%%%%%%%%%%%%%%%%%%%%%%%%%%
For the case of the monoenergetic signals coming from loop-induced
dark matter annihilations the background at these higher photon
energies ($E_{\gamma} \gappeq 100\GeV$) is substantially reduced
over the continuous differential signal considered just above. For
this case we include the ``low'' background of~(\ref{bkgrnd})
and~(\ref{lownorm}) plus the additional higher energy source
of~(\ref{ACT}). Together these provide a very small rate,
particularly when integrated over the narrow window
$E_{\gamma\gamma} \pm \sigma_{E}^{\rm ACT}$ or $E_{\gamma Z} \pm
\sigma_{E}^{\rm ACT}$ relevant for these monochromatic signals. For
example, if we integrate the background rate for a $1\sigma$ region
about $E=200\GeV$ we get $\order(10^{-2})$ events per m$^2$-year for
the low background and $\order(1)$ events per m$^2$-year for the
higher background which includes~(\ref{ACT}). Despite this pleasant
fact, the signal rates are very small as well. Typical annihilation
rates through these loop-induced processes are roughly~$10^{-3}$
to~$10^{-4}$ times those for the tree-level processes. For the
NFW~profile with density rescaling typical event rates in our model
set are $\sim 0.02$ photons/m$^2$-year.\footnote{With the rather
large effective area for a typical ACT experiment this small flux
can still produce a sizable photon count. Using the value of $A_{\rm
eff}$ in Table~\ref{gamexp}, and assuming 1000 hours of data-taking
per year with 75\% of $A_{\rm eff}$, gives a count rate of $\sim 50$
gamma ray photons per year.} Without rescaling this increases to
$\sim 10$ photons/m$^2$-year. Typical rates for other halo profiles
can be found multiplying these numbers by the ratios of
$\oline{J}(10^{-5}\,{\rm sr})$ values found in
Table~\ref{halomodel}.

Given the relatively low background rate we will require at least
$N_{\gamma}=10$ photons collected within $\pm \sigma_E^{\rm ACT}$ of
the expected signal(s) at $E_{\gamma\gamma}$ and $E_{\gamma Z}$. To
say two signals can be separated we continue to require that {\em
both} can first be conclusively detected. We therefore demand that
$N_i>m\sqrt{N_i^{\rm bkg}}$ where $i=\gamma \gamma$ or $\gamma\,Z$,
the background is the ``low'' background plus the ACT-contribution
of~(\ref{ACT}), and we will require a significance of $m=5$.
However, the limiting factor in our ability to distinguish models in
any given pair will not be signal significance but energy
resolution.

For a particular model with a given $m_{\chi}$ value the energies of
the two possible annihilation lines are given by~(\ref{lines}). For
an energy resolution of 15\% these two lines can be resolved only
for cases in which $m_{\chi} \lappeq 105 \GeV$. As this criteria is
met by only a handful of the models in our set we will therefore
consider a single line whose peak is at the average of the two
energies in~(\ref{lines}): $\oline{E} = (E_{\gamma \gamma} +
E_{\gamma Z})/2$.
We will then define two models to be distinguished if the two values
$\oline{E}^A$ and $\oline{E}^B$ are separated by $n\,\sigma_E$,
where $\sigma_E/E = 0.15$, for a particular requirement on the
significance~$n$.
For these degenerate model pairs the LSP masses $m_{\chi}$ for the
two models are often very similar -- implying that two values of
$\oline{E}$ are typically separated by no more than one standard
deviation for an energy resolution of 15\%. This makes separating
models extremely challenging in this arena. For example, let us take
the case of the NFW halo profile with adiabatic compression,
rescaling the local halo density by the ratio $r_{\chi}$. If we
consider an integrated exposure of 500~m$^2$-years at our generic
ACT (equivalent to about 200~hours of data taking at 75\% of the
$A_{\rm eff}$ value in Table~\ref{gamexp}) then in 161~of the
276~pairs both of the models would produce detectable monochromatic
signals above the ``high'' background of~(\ref{ACT}). Yet in only~67
pairs were the two central values $\oline{E}$ resolvable at the
$1\sigma$ level. Requiring $2\sigma$ resolution reduced this number
to only 23~pairs. If an energy resolution comparable to the GLAST
experiment of 10\% could be achieved, these numbers would roughly
double. A summary of what is possible for varying levels of
integrated exposure at a generic ACT is given in Table~\ref{mono}.
As the table suggests, the efficacy of ACTs in distinguishing
between models using monochromatic signals quickly saturates --
those model pairs with sufficiently large $m_{\chi}$ mass
differences as to be separated at a given significance level achieve
$S/\sqrt{B} \geq 5$ within 100-200 m$^2$-years of exposure.

%---------------- Monochromatic Table --------------------
\begin{table}[h]
\centering
\begin{tabular}{lccc|ccc}
ACT Exposure & \multicolumn{3}{c}{Rescaled} & \multicolumn{3}{c}{Not Rescaled} \\
 \hline
 & $1\sigma$ & $2\sigma$ & $3\sigma$ & $1\sigma$ & $2\sigma$ & $3\sigma$ \\
100 m$^2$-years & 51 & 14 & 3 & 66 & 23 & 10 \\
500 m$^2$-years & 67 & 23 & 10 & 95 & 25 & 11 \\
1000 m$^2$-years & 68 & 23 & 10 & 109 & 33 & 12 \\
\end{tabular}
%\end{center}
%
{\caption{\label{mono}\footnotesize {\bf Distinguishing pairs using
monochromatic signals}. The number of pairs that can be
differentiated for varying amounts of integrated exposure are given
for the generic ACT of Table~\ref{gamexp}. An NFW halo profile with
adiabatic compression has been assumed. In all cases we use the
background estimation of~(\ref{ACT}) with ${\rm signal}/\sqrt{\rm
bkgrnd} \geq 5$. The two lines $\oline{E}^A$ and $\oline{E}^B$ are
required to be separated by $n \, \sigma_E$, where $\sigma_E/E =
0.15$ and we consider $n=1$, 2~and~3. }}
\end{table}
%------------------------- END OF THE TABLE ---------------------

%%%% New paragraph on halo models %%%%
Returning to the issue of uncertainties in the halo model, it is
reasonable to ask whether the appearance of an indirect signal for
dark matter (whether it be a monochromatic line signal or the
overall integrated gamma ray flux from the galactic center) can tell
us anything about the supersymmetric model given the wide range of
$\oline{J}(\Delta \Omega)$ values listed in Table~\ref{halomodel}.
As mentioned earlier we believe the answer would be ``no'' in the
absence of other data. But the assumption in this paper has been
that we have two candidate models that have been constructed from
the LHC data. For a given halo profile these models make two
concrete predictions for the signals considered in this section. We
have been asking whether the experiments considered here have the
inherent power to resolve these predictions. The result of this
section is that, given enough statistics, the answer is typically
yes. It is important to note that the normalizations parameterized
by $\oline{J}(\Delta \Omega)$ differ from one another by two orders
of magnitude. Only two pairs of models in our study gave predictions
that differed by this much or more for the integrated gamma ray flux
-- and no pairs differed by this amount for the monochromatic
predictions. In fact, 252~of the~276 pairs of models gave
predictions for the integrated gamma ray fluxes that differed by
less than an order of magnitude. If we assume that the choices in
Table~\ref{halomodel} should be treated as a discrete set of
possibilities then the size of any observed signal at GLAST or some
future ACT experiment will likely pick out only {\em one} halo
profile as reasonable if a fit is to be made to our post-LHC
degenerate models.

Alternatively, if we treat the quantity $\oline{J}(\Delta \Omega)$
as an undetermined free parameter we can ask how well we would need
to know the value of this parameter {\em a priori} to be able to
distinguish the pairs of models in our list. To investigate this
question we computed the total integrated gamma ray flux from the
galactic center between the energy ranges of 1~GeV to 200~GeV, using
the $\oline{J}(\Delta \Omega)$ value for GLAST for the NFW profile
with adiabatic compression. We then converted this flux into a
numerical count for our model pairs assuming no halo rescaling and
3~m$^2$-years of exposure. We then asked how many model pairs could
be separated at the~$3\sigma$ level assuming the value of
$\oline{J}$ was uncertain by an amount
\begin{equation} \(\delta \oline{J} \)_{\rm theor}
= \epsilon \times \oline{J}(\Delta \Omega)\, , \label{epsilon2}
\end{equation}
analogous to the consideration of nuclear matrix element
uncertainties in Section~\ref{sec:direct}. If we allow a 5\% error
in the input value of $\oline{J}(\Delta \Omega)$ then we can
separate 152, 102 and 22 model pairs out of 276 for the cases of no
background, low background, and high background, respectively. These
numbers should be compared with the entries in the corresponding
column of Table~\ref{gamma1}. The number of separable model pairs
drops steadily, reaching zero for $\epsilon \simeq 0.32$ (low
background) and $\epsilon \simeq 0.26$ (high background). Clearly,
then, the theoretical modeling that goes into producing the values
in Table~\ref{halomodel} will need to be accurate to the 5-10\%
range to truly be able to separate post-LHC candidate supersymmetric
models from gamma-ray observations alone.

\section{Conclusions}

If supersymmetry is relevant to the physics of the electroweak scale
then it is very likely to be discovered in the near future at the
LHC. Yet if the results of~\cite{ArkaniHamed:2005px} are indicative
of general supersymmetric theories then it is also likely that more
than one supersymmetric model (or low energy parameter set) will be
a reasonable fit to the ensemble of LHC measurements that will be
made. In such a scenario the challenge to the high energy community
will be to find orthogonal information that will be effective in
breaking these degeneracies. If the lightest supersymmetric particle
is stable then it is reasonable to imagine that it will be detected
at future dark matter experiments. It is therefore instructive to
consider how well measurements in this area serve to provide the
needed orthogonal data.

In the current work we have considered only a subset of the
experimental data that might be available over the next decade in
the area of direct and indirect detection of dark matter
neutralinos. However taken together these signals are sufficient to
separate a large number of the degenerate pairs of Arkani-Hamed et
al., even when rather conservative assumptions are made.
Unfortunately this statement comes with a large caveat: the ability
to distinguish between models will depend on certain theoretical
inputs being better understood. For example, if superpartners exist
at the electroweak scale and the LSP is stable then it is almost
certain that a one-ton liquid xenon detector will eventually see a
neutralino-recoil signal. But determining whether that signal was
consistent with only one of two competing SUSY models (our original
thought experiment) will be impossible without much better knowledge
of the nuclear matrix elements that appear in the cross-section
calculation. In this work we have made the assumption that such
knowledge can indeed be obtained within the ten year time horizon we
imagine between now and any future~ILC experiment.

%%%%% New paragraph %%%%%
Another theoretical uncertainty is the imperfect knowledge of the
dark matter halo in our galaxy -- both the local density near the
earth as well as the dependence of the halo density as a function of
position from the galactic center. These quantities are fundamental
inputs to any theoretical prediction, and as we have seen they can
vary over larger ranges than other (in principle tractable)
theoretical uncertainties. The existence of this ambiguity is an
inherent property of cosmological observations; experiments will
measure count rates which require both a particle physics model
($\lang \sigma v \rang$) and an astrophysics model
($\oline{J}(\Delta \Omega)$) as inputs. Our theoretical control over
the former -- particularly in a post-LHC world -- will likely be
much better than over the latter. But this need not imply that dark
matter signals have nothing to tell us about supersymmetric models.
It is likely true that {\em on their own} such signals will be
unable to establish any particular SUSY model as the correct one.
But if any future dark matter signal proves to be consistent with
what will be known at the LHC, for reasonable input assumptions,
then this will be powerful (if indirect) evidence in favor of those
assumptions. If, in addition, the observed signal is consistent with
only one of a pair of predictions {\em based on those same
assumptions} then we can reasonably claim that the dark matter
signal in question has resolved a potential degeneracy. Of course if
any future signal cannot be made consistent with a post-LHC
candidate model, or is consistent only with unrealistic input
assumptions, then no such resolving power exists.
%%%%%%%

With this philosophy in mind, we summarize the effect of all
experiments considered in this work in Table~\ref{final}, where the
number of pairs that are distinguishable are listed under three sets
of assumptions which we label as Conservative, Moderate and
Optimistic:
\begin{description}
\item [Conservative]
All rates are rescaled by the calculated thermal relic density via
the quantity $r_{\chi} = {\rm Min}(1,\, \Ochi/0.025)$. We require at
least 100~recoil events in direct detection experiments, at least
100~photons over the range $1\GeV \leq E_{\gamma} \leq 200\GeV$ for
gamma ray spectral observations, and at least 10 photons for
monochromatic gamma ray observations. For direct detection
experiments we demand $5\sigma$ separation of models, including an
additional factor $f=0.2$ in the error. We assume 100~kg-years of
exposure in germanium and 1 ton-year of exposure in xenon. For gamma
ray experiments we assume the NFW halo profile with adiabatic
compression and the ``high'' background for both GLAST and our
generic ACT. We imagine 5~m$^2$-years of exposure for GLAST and
1000~m$^2$-years of exposure for the ACT experiment. Signals are
required to be separable by $5\sigma$ ($2\sigma$) and be above
background by $2\sigma$ ($5\sigma$) for the continuous gamma ray
observations (monochromatic lines).
\item [Moderate] Same as the conservative case, but we now assume an
essentially background-free environment at the direct detection
experiments. Consequently, we set $f=0$ in the error estimate for
direct detection experiments and require only 10~recoil events. For
the gamma ray observables we assume the ``low'' background rate and
take 2500~m$^2$-years of exposure for the ACT experiment.
\item [Optimistic] Same as the moderate case, but now all rates are
computed without rescaling the neutralino density $\rho_{\chi}$. We
further assume that germanium experiments will eventually reach 1
ton-year of exposure, liquid xenon experiments will reach
5~ton-years, and that the ACT experiment will reach
10000~m$^2$-years of exposure.
\end{description}
All assumptions assume perfect ability to predict theoretical
interaction rates from a given supersymmetric model. With the most
conservative assumptions we achieve a 37\% success rate at
separating all model pairs, and 44\% for the pairs that we denoted
as ``physical'' in Section~\ref{sec:constraints}. With the only
slightly less conservative estimates we calculate these numbers to
nearly double to 67\% and 81\%, respectively. To achieve a much
larger degree of separation between the models (the optimistic
scenario) one must assume that the energy density of the stable LSP
is given $\rho_{\chi} = 0.3 \GeV/{\rm cm}^3$ regardless of the
predicted thermal relic abundance. We note, however, that
approximately 60\% of the models that have no charged states with
masses below 240~GeV can be distinguished, even with the moderate
scenario. These are precisely the models which a $\sqrt{s} =
500\GeV$ linear collider will have difficulty separating -- and such
a linear collider is unlikely to exist until {\em after} the
measurements considered in this paper have been made.

%---------------------- Summary Table -------------------------
\begin{table}[t]
\begin{center}
\begin{tabular}{|l||c|c|c|}
\multicolumn{1}{c}{}
 & \multicolumn{1}{c}{Conservative}
 & \multicolumn{1}{c}{Moderate}  & \multicolumn{1}{c}{Optimistic} \\ \hline
All Pairs &  &  & \\
\mbox{         } Direct detection, xenon & 48 & 112 & 224\\
\mbox{         } Direct detection, germanium & 4 & 14 & 147\\
\mbox{         } Gamma rays, continuum & 56 & 115 & 158\\
\mbox{         } Gamma rays, monochromatic & 23 & 34 & 36\\ \hline
All Pairs, All Signals  & 101 & 186 & 245 \\ \hline
\mbox{         } Physical Pairs Only  & 34 & 55 & 77\\
\mbox{         } ILC Inseparable Only & 32 & 62 & 81 \\ \hline
\end{tabular}
\end{center}
{\caption{\label{final}\footnotesize {\bf Final distinguishability
analysis}. Final number of pairs from the original set of~276 pairs
which can be distinguished using all experimental data considered in
this work. In the upper section of the table we break down the total
by signal. Note that many pairs can be distinguished by more than
one set of observations. The set of 77~physical pairs were defined
in Section~\ref{sec:constraints}. The set denoted ``ILC
Inseparable'' are the 103~pairs for which neither model had a
charged superpartner below $240\GeV$ in mass. For the set of
assumptions which define the Conservative, Moderate and Optimistic
scenarios, see text.}}
\end{table}
%------------------------- END OF THE TABLE ---------------------

To truly realize the potential of future dark matter searches and
achieve the results suggested in Table~\ref{final} it will be
necessary to do much better on taming the uncertainties associated
with theoretical inputs, particularly for the galactic halo model.
In the absence of significant improvement on this front it may be
necessary to use multiple observations to constrain the theoretical
inputs. In this paper we have considered only a handful of possible
observations, but many more indirect signatures of relic neutralino
dark matter could have been included in this analysis, including
spin-dependent interactions on targets such as CF$_3$I at
COUPP~\cite{Bertone:2007xj} (or even on germanium targets),
annihilation of neutralinos into antimatter or neutrinos, and
various properties such as the absolute or relative intensities of
monochromatic annihilation signals. We chose the observables
considered in this paper because they appear to have the best
inherent resolving power for distinguishing between LHC
degeneracies, modulo theoretical uncertainties. For any particular
model all of these signatures are related to one another in a
calculable way. It is possible, therefore, that given an ensemble of
cosmological observations one candidate model can be favored over
another even without a full understanding of the theoretical inputs
behind the halo profile. Further research in this direction is
therefore extremely important.

We are encouraged by the recent results of Bernal et
al.~\cite{Bernal:2008zk} in which it was demonstrated that these
same experiments will likely make very good measurements of the
properties of the LSP -- using similar assumptions about the
galactic halo, background rates, and methodology. Clearly the high
energy community is beginning to imagine the day when dark matter
experiments are making {\em measurements} and not simply setting
limits. For example, recent work suggests it may be possible to
estimate the mass of the recoiling WIMP particle in direct detection
experiments~\cite{Hooper:2006wv}. This information will surely be of
utmost importance in breaking degeneracies, given the likely ability
of the LHC to measure the mass of the invisible, stable
LSP~\cite{Cheng:2007xv}. In this study we have chosen to follow the
spirit of~\cite{ArkaniHamed:2005px} and have not considered this
additional information. We anticipate that the results contained in
this work can be further refined and sharpened by those with greater
expertise in the relevant experimental conditions.
We also believe it would be useful to consider the broader ensemble
of dark matter experiments and observations available in the near
future. Multiple observations only add to the power to separate
models and provide important cross-checks on input assumptions. Just
to give one example, it has recently been suggested that the
excellent angular resolution of the GLAST experiment may make it
possible to measure properties of the galactic halo directly from
the data -- particularly if the mass of the LSP is known or can be
inferred from other data~\cite{Dodelson:2007gd}. Given the high
degree of sensitivity of the numbers in Table~\ref{final} to such
assumptions it will no doubt be necessary to observe multiple
signals which can be reconciled with a common, consistent model if
dark matter data is to be persuasive in solving the LHC inverse
problem.

\section*{Acknowledgments}
It is a pleasure to thank Jesse~Thaler for providing us with the
list of input model parameters from the study in
Reference~\cite{ArkaniHamed:2005px}, as well as answering several
questions on their interpretation. We are grateful to Gabe
Shaughnessy and Steve Martin for pointing out the need to consider
the uncertainty associated with theoretical matrix elements in
neutralino-nucleon interactions. We also would like to thank
Haim~Goldberg and Dan~Feldman for useful discussions, as well
Elena~Aprile, Richard~Gaitskell and Dan~Hooper for assistance with
various experimental questions. This work was supported by National
Science Foundation Grant PHY-0653587.

%\pagebreak

\end{document}